\documentclass[twocolumn,showpacs,aps,epsfig,nofootinbib]{revtex4}

\usepackage{graphicx}
\usepackage{epstopdf}
\usepackage{latexsym}
\usepackage{amssymb}
\usepackage{color}

\usepackage[center]{subfigure}

\begin{document}

 \newcommand{\bq}{\begin{equation}}
 \newcommand{\eq}{\end{equation}}
 \newcommand{\bqn}{\begin{eqnarray}}
 \newcommand{\eqn}{\end{eqnarray}}
 \newcommand{\nb}{\nonumber}
 \newcommand{\lb}{\label}
\newcommand{\PRL}{Phys. Rev. Lett.}
\newcommand{\PL}{Phys. Lett.}
\newcommand{\PR}{Phys. Rev.}
\newcommand{\CQG}{Class. Quantum Grav.}

\title{High-dimensional Lifshitz-type  spacetimes, universal horizons  and black  holes  in  Ho\v{r}ava-Lifshitz gravity}

\author{Kai Lin ${}^{a, b}$}
\email{lk314159@hotmail.com}

 \author{Fu-Wen Shu $^{ c}$}
\email{shufuwen@ncu.edu.cn}

\author{Anzhong Wang $^{a, d}$\footnote{Corresponding author}}
\email{anzhong_wang@baylor.edu}

\author{Qiang Wu ${}^{a}$}
\email{wuq@zjut.edu.cn}

\affiliation{$^{a}$ Institute  for Advanced Physics $\&$ Mathematics,
Zhejiang University of
Technology, Hangzhou 310032,  China\\
${}^{b}$ Instituto de F\'isica, Universidade de S\~ao Paulo, CP 66318, 05315-970, S\~ao Paulo, Brazil \\
${}^{c}$ Center for Relativistic Astrophysics and High Energy Physics, Nanchang University, Nanchang 330031, China\\
$^{d}$ GCAP-CASPER, Physics Department, Baylor University, Waco, TX 76798-7316, USA}

\date{\today}

\begin{abstract}

In this paper, we present all $[(d+1)+1]$-dimensional static  diagonal vacuum solutions of the  non-projectable Ho\v{r}ava-Lifshitz 
gravity in the IR limit, and show that they give rise to very rich Lifshitz-type structures, depending on the choice of the free parameters 
of the solutions. These include  the Lifshitz spacetimes with or without hyperscaling violation, Lifshitz solitons, and black holes.   
Remarkably,  even the  theory breaks explicitly  the Lorentz symmetry  and allows generically instantaneous propagations, universal 
horizons  still exist, which  serve as one-way membranes for signals with any large velocities. In particular, particles even with
infinitely large velocities would just move around on these boundaries and  cannot escape to infinity. Another remarkable  feature 
appearing in the Lifshitz-type spacetimes is that the dynamical exponent $z$ can take its values only in the  ranges $1 \le z < 2$ for 
$d \ge 3$ and $1 \le z <\infty$  for $d = 2$, due to the stability and ghost-free conditions of the theory.

\end{abstract}


\maketitle

\section{ Introduction  }
\renewcommand{\theequation}{1.\arabic{equation}} \setcounter{equation}{0}

Lifshitz space-time has been extensively studied in the content of non-relativistic gauge/gravity duality \cite{Mann,HKW}, after the seminal work of
\cite{KLM}, which argued that nonrelativistic quantum field theories (QFTs) that describe multicritical points in certain magnetic materials and liquid crystals \cite{Sachdev}
may be dual to  certain nonrelativistic  gravitational theories in such a space-time background.

One of the remarkable feature of the Lifshitz space-time is its anisotropic scaling between space and time,
\bq
\lb{1.1}
 t \rightarrow b^{z} t, \;\;\; { x^i} \rightarrow b { x^i},
\eq
on a hypersurface $r = $ Constant, on which the nonrelativistic QFTs live, where $z$ denotes the dynamical critical exponent, and in the relativistic scaling we have
$z_{GR} = 1$.  $x^i$ denote the spatial  coordinates tangential to the surfaces $t= $ Constant.

It is interesting to note that the anisotropic scaling (\ref{1.1}) can be realized in two different levels. In Level one, the underlying  theory itself is still relativistic-scaling invariant,
but the space-time has the anisotropic scaling. This was precisely the case studied in \cite{KLM,Mann}, where the theories of gravity is still of general covariance, but
the metric of the space-time has the above anisotropic scaling. This is possible  only  when some matter fields are introduced to create a preferred direction, so that the
anisotropic scaling (\ref{1.1}) can be realized. In \cite{KLM}, this was done by two p-form gauge fields with $p = 1, 2$, and was soon generalized to different  cases   \cite{Mann}.

In Level two, not only the space-time has the above anisotropic scaling, but also the theory itself. In fact,   starting with the anisotropic scaling (\ref{1.1}), Ho\v{r}ava
recently  constructed a  theory of quantum gravity at a Lifshitz fixed point,  the so-called  Ho\v{r}ava-Lifshitz (HL) theory \cite{Horava},
which is  power-counting renormalizable, and  lately has attracted lots  of attention, due to its remarkable features when applied to cosmology and
astrophysics \cite{reviews}. Power-counting renomalizability requires $z \ge D$, where $D$ denotes the number of spatial dimensions of the theory.
Since the anisotropic scaling (\ref{1.1}) is built in by construction in the HL gravity, it is natural to expect that the HL gravity provides a minimal holographic dual for
non-relativistic Lifshitz-type field theories with the anisotropic scaling. Indeed,  this was  first showed in \cite{GHMT}
that the Lifshitz spacetime,
\bq
\lb{1.2}
ds^2 = - \left(\frac{r}{\ell}\right)^{2z} dt^2 + \left(\frac{r}{\ell}\right)^{2}dx + \left(\frac{\ell}{r}\right)^{2} dr^2,
\eq
is a vacuum solution of the HL gravity in (2+1) dimensions, and that   the full structure of the $z=2$ anisotropic Weyl anomaly can be reproduced  in dual field theories,
while  its minimal relativistic gravity counterpart yields only one of two independent central charges in the anomaly.

Recently, we  studied the HL gravity in (2+1) dimensions in detail \cite{SLWW}, and found further evidence to support the above speculations. In particular, we found all
the static (diagonal) solutions of the HL gravity in (2+1) dimensions, and showed that they give rise to      very rich space-time structures: the corresponding spacetimes
can represent the generalized Banados, Teitelboim and Zanelli (BTZ) black holes \cite{BTZ},  the Lifshitz space-times,  or Lifshitz solitons \cite{LSolitons}, in which the spacetimes are free of any kind
of space-time singularities, depending on the choices of the free parameters of the solutions. Some space-times are not complete, and extensions beyond certain horizons
are needed. In addition, it was shown recently that the Lifshitz space-time (\ref{1.2}) is not only a solution of the HL gravity in the IR, but also a solution of the full theory, that is, 
even high-order operators are all included \cite{WYTWDC}. The only effects of these high-order operators are to shift    $z$ from one value to another, as longer as
the spacetime  itself is concerned.  

In this paper, we shall generalize our above studies to any dimensions, and obtain all the static (diagonal) solutions of the vacuum HL gravity explicitly. 
With these exact vacuum solutions,we believe  that the studies of the non-relativistic Lifshitz-type gauge/gravity duality will be simplified considerably, as
so far most of such studies are numerical \cite{Mann,GR10,LSolitons,LBHs,OTU}.  After studying each of
these solutions in detail, we find that, similar to the (2+1) case, Lifshitz  space-times and solitons     can be all found in
these solutions. Remarkably, the Lifshitz  space-times with hyperscaling violation \cite{GR10,OTU},
\bq
\lb{1.3}
ds^2 = r^{- \frac{2(d-\theta)}{d}}\left(- r^{-2(z-1)} dt^2 +  dr^2 + d\vec{x}^2\right),
\eq
can be also realized in the HL gravity as a vacuum solutions of the theory.  

Moreover, some of the solutions to be presented in this paper also represent black holes, although the HL gravity explicitly breaks Lorentz symmetry and  allows 
in principle propagations with any large velocities \cite{Horava,reviews}.  This follows the recent discovery of the  existence of the universal horizons in the  khrononmetric 
theory of gravity \cite{BS11}, in which the khronon $\phi$ naturally defines a timelike foliations, parametrized by $\phi\left(x^{\mu}\right) = $ Constant. Among
these leaves,  there may exist a   surface at which $\phi$ diverges, while physically nothing
singular happens there, including  the metric and the space-time. Given that $\phi$ defines an absolute time, any object crossing this surface from the interior would necessarily also move back in
absolute time, which is something forbidden by the definition of the causality in the theory. Thus, even particles with superluminal velocities cannot penetrate this surface, once they are
trapped inside it. In particular, particles even with infinitely large velocities     would just move around on these
boundaries and  cannot escape to infinity.  For more details, we refer readers to \cite{BS11,UHs,LACW}.

The rest of the paper is organized as follows: In Section II, we give a brief introduction
to the (d+2)-dimensional HL gravity without the projectablity condition, while in Section III we first write down the corresponding field equations for static vacuum spacetimes,
and then solve them for particular cases.  In Section IV, we first obtain all the rest of the static (diagonal) vacuum (d+2)-dimensional solutions of the HL theory, and then study
each of such solutions in detail. In Section V, following \cite{LACW,LGSW} we study the black hole structures of solutions presented in Section III, and show explicitly that universal horizons
exist in some of these solutions. Finally, in Section VI we present our main conclusions and provide some discussing remarks.

\section{ Non-projectable HL theory  in $D$ dimensions  }
\renewcommand{\theequation}{2.\arabic{equation}} \setcounter{equation}{0}

In this paper, we shall take the Arnowitt-Deser-Misner (ADM) variables \cite{ADM},
 \bq
\lb{2.0} \left(N, N_i, g_{ij}\right),\; (i,\; j = 1, 2,
\cdot\cdot\cdot, d+1),
 \eq
as the  fundamental  ones,  which are all functions of both $t$ and $x^i$, as in this paper we shall work in the version of the HL
gravity without the projectability condition \cite{Horava,reviews}. Then, the general action of the HL theory
in (d+2)-dimensions is given by
 \bqn
  \lb{2.1}
S &=& \zeta^2\int dt d^{d+1}x N \sqrt{g} \Big({\cal{L}}_{K} -
{\cal{L}}_{{V}}   +{\zeta^{-2}} {\cal{L}}_{M} \Big), ~~~~
 \eqn
where $  g={\rm det}(g_{ij})$, $\zeta^2 = {1}/{(16\pi G)}$, and
 \bqn
\lb{2.2a}
&& {\cal{L}}_{K} = K_{ij}K^{ij} -   \lambda K^{2},\nb\\
&& K_{ij} =  \frac{1}{2N}\left(- \dot{g}_{ij} + \nabla_{i}N_{j} +
\nabla_{j}N_{i}\right).
 \eqn
Here $\lambda$ is a dimensionless
coupling constant, and $\nabla_i$ denotes the covariant derivative with respect to $g_{ij}$.  ${\cal{L}}_{{M}}$ is the Lagrangian of matter
fields. The potential $ {\cal{L}}_{V}$ is constructed from $R_{ij},\; a_i$ and $\nabla_i$, and formally can be written  in the form,
 \bqn
  \lb{2.2}
 {\cal{L}}_{V}   & = &   \gamma_{0}\zeta^{2}  + \gamma_1R + \beta  a_{i}a^{i}
+ {{\cal{L}}}_{V}^{z>2}\left(R_{ij}, a_i, \nabla_i\right), ~~~
 \eqn
where   ${{\cal{L}}}_{V}^{z>2}$ denotes the part that  includes
all  higher-order operators \cite{ZWWS}. Power-counting renormalizability condition requires $z \ge (d+1)$  \cite{Horava,reviews}.  $R_{ij}$ denotes the Ricci tensor made of $g_{ij}$, and
 \bqn
 \lb{2.3}
a_i &\equiv& \frac{N_{,i}}{N},\;\;\; a_{ij} \equiv \nabla_{i}a_j.
 \eqn

In the infrared (IR) limit, the higher-order operators are suppressed by $M_*^{2-n}$,  so we can safely
set them to zero,
\bq
\lb{IR}
{{\cal{L}}}_{V}^{z>2}\left(R_{ij}, a_i, \nabla_i\right) = 0,
\eq
where $M_{*} \equiv 1/\sqrt{8\pi G}$ and $n$ denotes the order of the operator. In this paper, we shall consider only the IR limit, so that
Eq.(\ref{IR}) is always true.

 \subsection{Field Equations in IR Limit}

Variation of the action (\ref{2.1}) with respect to the lapse
function $N$ yields the Hamiltonian constraint
\bqn
\label{hami}
 {\cal{L}}_K + {\cal{L}}_V^R + F_V= 8\pi G J^t,\;\;
\eqn
where
\bqn
J^t&=& 2\frac{\delta (N\mathcal{L}_M)}{\delta N},\nb\\
{\cal{L}}_V^R &=& \gamma_0 \zeta^2+\gamma_1R,\nb\\
F_V&=&-\beta\left(2a^i_i+a_ia^i\right).
 \eqn

Variation with respect to the shift vector $N_i$ yields the momentum constraint
\bqn\lb{momen}
\nabla_j \pi^{ij}=8\pi G J^i,
\eqn
where
\bq
\pi^{ij}\equiv  -K^{ij}+\lambda K g^{ij},\ \ \ J^{i}\equiv -  \frac{\delta \left(N {\cal{L}}_M\right) }{\delta N_i}.
\eq

The dynamical equations are obtained by varying the action with respect to $g_{ij}$, and are given by
\bqn \label{dyn}
\frac{1}{\sqrt{g}N} \frac{\partial}{\partial t}\left(\sqrt{g} \pi^{ij}\right)+2(K^{ik}K^j_k-\lambda K K^{ij})\nb\\
-\frac12g_{ij}{\cal{L}}_K+\frac{1}{N}\nabla_k (\pi^{ik}N^j+\pi^{kj}N^i-\pi^{ij}N^k)\nb\\
-F^{ij}-F^{ij}_a=8\pi G \tau^{ij},\;\;\;\;\;\;
\eqn
where
 \bqn
\lb{tauij}
\tau^{ij}&\equiv&\frac{2}{\sqrt{g}N} \frac{\delta(\sqrt{g}N{\cal{L}}_M)}{\delta g_{ij}}, \nb\\
F^{ij}&\equiv&\frac{1}{\sqrt{g}N}\frac{\delta (-\sqrt{g}N
{\cal{L}}_V^R)}{\delta g_{ij}}\nb\\
            &=& -\Lambda g^{ij}+\gamma_1\left(R^{ij}-\frac{1}{2}Rg^{ij}\right)\nb\\
            &&+\frac{\gamma_1}{N}\left(g^{ij}\nabla^2N-\nabla^i\nabla^jN\right),\nb\\
F^{ij}_a&\equiv&\frac{1}{\sqrt{g}N}\frac{\delta (-\sqrt{g}N
{\cal{L}}_V^a)}{\delta g_{ij}}\nb\\
               &=& \beta \left(a^ia^j-\frac{1}{2}g^{ij}a^ka_k\right),
 \eqn
with $ {\cal{L}}_V^a \equiv \beta  a_{i}a^{i}$.

In addition, the matter components $(J^t, J^i,  \tau^{ij})$ satisfy the conservation laws of energy and momentum,
\bqn
\label{energy conservation}
&&\int d^3x \sqrt{g} N \bigg[\dot{g}_{ij}\tau^{ij}-\frac{1}{\sqrt{g}}\partial_t (\sqrt{g} J^t)\nb\\
&& ~~~~ +\frac{2 N_i}{\sqrt{g} N}\partial_t (\sqrt{g} J^i)\bigg]=0,\\
\label{mom conservation}
&& \frac{1}{N}\nabla^i(N\tau_{ik})-\frac{1}{\sqrt{g} N} \partial_t (\sqrt{g} J_k) -\frac{J^t}{2N}\nabla_kN\nb\\
&&  ~~~~ -\frac{N_k}{N} \nabla_i J^i-\frac{J^i}{N}(\nabla_i N_k-\nabla_kN_i) =0.
\eqn

\subsection{Stability and Ghost-free Conditions}

When $\gamma_0 = 0$, the above HL theory admits the Minkowski space-time
\bq
\lb{2.15}
\left(\bar{N}, \bar{N}_i, \bar{g}_{ij}\right) = \left(1, 0, \delta_{ij}\right),
\eq
as a solution of the theory. Then, its linear perturbations reveals that the theory has two modes \cite{GHMT}, one represents the spin-2 massless
gravitons with a dispersion relation,
\bq
\lb{2.16}
\omega_{T}^2 = -\gamma_1 k^2,
\eq
and the other represents the scalar mode with
\bq
\lb{2.17}
\omega_{S}^2 = -\frac{\gamma_1(\lambda -1)}{(d+1)\lambda - 1}\left[d\left(\frac{\gamma_1}{\beta} -1\right)+1\right] k^2.
\eq
The stability conditions of these modes requires
\bq
\lb{2.18}
\omega_{T}^2 > 0,\;\;\;
\omega_{S}^2 > 0,
\eq
for any given $k$.

On the other hand, the kinetic term of the scalar mode is proportional to $(\lambda -1)/[(d+1)\lambda - 1]$ \cite{GHMT}, so the
ghost-free condition requires
\bq
\lb{2.19}
\frac{\lambda -1}{(d+1)\lambda - 1} \ge 0,
\eq
which is equivalent to
\bq
\lb{2.20a}
i)\; \lambda \ge 1,\;\;\; {\mbox{or}} \;\;\;\; ii)\; \lambda \le \frac{1}{d+1}.
\eq
Then, Eq.(\ref{2.18}) implies that \footnote{It is interesting to note that in (2+1)-dimensions, the spin-2 gravitons do not exist, so the
coupling constant $\gamma_1$ is free, while  $\beta$ is required to be negative, $\beta < 0$ \cite{SLWW}.}
\bq
\lb{2.20b}
\gamma_1 < 0, \;\;\;
 \frac{d\gamma_1}{d-1} < \beta < 0.
\eq

\section{Static vacuum solutions}

\renewcommand{\theequation}{3.\arabic{equation}} \setcounter{equation}{0}

In this paper, we consider static  spacetimes  given by,
\bqn
\lb{3.2}
N &=& r^z f(r),\;\;\; N^i = 0,
\nb\\
g_{ij}dx^idx^j  &=& \frac{g^2(r)}{r^2}dr^2 +  r^2d\vec{x}^2,
\eqn
in the coordinates $\left(t,  x^A, r\right), (A=1, 2, \cdots, d)$, where $d\vec{x}^2 \equiv \delta_{AB} dx^Adx^B$. Note that in \cite{SLWW}, the case $d = 1$ was studied in detail.
So, in this paper we shall consider only the case where
$d \ge 2$.

Then, the  $(d+1)-$dimensional Ricci scalar
$R \; \left(\equiv g^{ij}R_{ij}\right)$ of the leaves $t = $ Constant  is given by
\bqn
\lb{Ricci}
R=   \frac{d}{g^{3}(r)}\left[2rg'(r) - (d+1)g(r)\right].
\eqn

On the other hand, since $N^i= 0$ and that the spacetimes are static, so we must have  $K_{ij} = 0$. Then,   the momentum
constraint (\ref{momen}) is satisfied identically.  The Hamiltonian constraint (\ref{hami}) and the $rr$-component  of the dynamical equations (\ref{dyn}) are non-trivial, while
the $AA$-component of the dynamical equations can be
derived from the Hamiltonian constraint and the $rr$ component. Therefore, similar to the (2+1)-dimensional case, there are only two independent
equations for two unknowns, $f(r)$ and $g(r)$, which can be cast in the forms,
\bqn
\lb{hami2}
&& \Lambda g^2-d\gamma_1W-\frac{1}{2}\beta W^2-\frac{1}{2}d(d-1)\gamma_1 = 0,\\
 \lb{dyn1}
&& \Lambda g^2-\beta \left[\left(\frac{rW}{g}\right)'+(d-1)W+\frac{W^2}{2}\right] \nb\\
&&~~~~~~~~ -d\gamma_1\left[\frac{d+1}{2}-r\frac{g'}{g}\right]=0,
 \eqn
where
\bq
\lb{ff}
W \equiv z + r\frac{f'}{f},\;\;\; \Lambda\equiv \frac{1}{2} \gamma_0\zeta^2.
\eq

From Eq.(\ref{hami2}), we obtain
\bq
\lb{Fpm}
W_{\pm}=\frac{s[1\pm r_*(r)]}{1-s},
\eq
where
\bqn
\lb{Fpm0}
s& \equiv& \frac{d\gamma_1}{d\gamma_1-\beta},\nb\\
r_*(r)& \equiv& \sqrt{1+(1-d)\frac{\beta}{d\gamma_1}+\frac{2\beta\Lambda}{d^2\gamma_1^2}g(r)^2}.
\eqn
Then, from the stability conditions (\ref{2.20b}) we find that
\bq
\lb{sCondt}
1 \le s <  \frac{d-1}{d-2},
\eq
where the equality holds only when $\beta = 0$, which is possible when $\lambda =1$, as can be seen from Eq.(\ref{2.17}).

Inserting the above expression  into Eq.(\ref{dyn1}), we obtain a master
equation for $r_*(r)$,
\bqn
\lb{hami3r}
(s-1)rr'_*+\Delta\left( r^2_*-r_s^2\right)
\left(r_* + \epsilon {\cal{D}}\right)=0,
\eqn
where $\epsilon = \pm 1$,  and
\bqn
\lb{pms}
r_{s}^2 &\equiv& 1 - \frac{(d-1)\beta}{d\gamma_1},\;\;\;
{\cal{D}} \equiv \frac{d\gamma_1 - (d-1)\beta}{d(\gamma_1 -\beta)}, \nb\\
\Delta &\equiv& \frac{d^2\gamma_1(\gamma_1 - \beta)}{(d\gamma_1-\beta)[d\gamma_1 - (d-1)\beta]}.
\eqn

Note that  Eq.(\ref{hami3r})   with $``-''$ sign can be always obtained from the one with $``+''$ sign,  by simply replacing $r_*$ by $-r_*$.
Therefore, although $r_*$  defined by Eq.(\ref{Fpm0})
is non-negative, we shall take the region $r_*<0$ as a natural extension, so that
in the following  we only need to consider   the case with $``+''$ sign.

From Eq.(\ref{Fpm0}) we find that,
\bqn
\lb{case2g}
g^2(r)=\frac{d^2\gamma_1^2}{2\beta\Lambda}\left(r_*^2-r_s^2\right),
\eqn
while from Eqs.(\ref{ff}) and (\ref{Fpm}), we obtain
\bqn
\lb{case2f}
\frac{df}{f}=\frac{s-z+zs+\epsilon sr_*}{1-s}\left(\frac{dr}{r}\right).
\eqn

Therefore, once the master equation (\ref{hami3r}) is solved for $r = r(r_*)$, substituting it into Eq.(\ref{case2f}) we can find $f(r_*)$. Then, in terms of $r_*$, the metric takes the form,
\bq
\lb{metricrstar}
ds^2 = - r^{2z}f^2dt^2 + \frac{g^{2}}{r^2}\left(\frac{dr}{dr_*}\right)^2 dr_*^2 + r^2d\vec{x}^2.
\eq

 In the rest of this section, we shall solve the above equations for some particular cases, and leave  the one with  $r_s^2 > 0$ to the next section.

 \subsection{Lifshitz Spacetime}

A particular solution of Eq.(\ref{hami3r}) is $r_*=-\epsilon {\cal D}$. Then we  obtain
\bqn
 \lb{case1fg}
g^2(r)&=&g_0^2,\;\;\;\;
f(r) = f_0 r^{\frac{s }{d+s(1-d)} - z},
\eqn
where in terms of $g_0$, the cosmological constant is given by,
\bqn
\lb{case1a}
\Lambda=\gamma_1\frac{(\beta-d\beta+d\gamma_1)(\gamma_1-d\beta+d\gamma_1)}{2g_0^2(\gamma_1-\beta)^2},
\eqn
with $f_0$ and $g_0$ being the integration constants. Then, the corresponding line element takes the form,
\bqn
 \lb{Lifshitz Vacuum}
ds^2&=& {L^2}\left\{-\left(\frac{r}{\ell}\right)^{2z}dt^2+
\left(\frac{\ell}{r}\right)^{2} dr^2  \right.\nb\\
&& ~~~~~~ \left. +
\left(\frac{r}{\ell}\right)^{2} d\vec{x}^2\right\},
\eqn
where   $f_0 \equiv L/\ell^z,\; g_0 \equiv L \ell$, and
\bq
\lb{zexp}
z = \frac{\gamma_1}{\gamma_1 - \beta},
\eq
which is independent of the space-time dimensions. On the other hand, from the stability and ghost-free condition (\ref{2.20b}), it can be shown that
\bq
\lb{3.18}
1 \le z < \frac{d-1}{d-2}.
\eq
Note that the above holds only for $d \ge 2$. In particular, we have
\bq
\lb{3.19}
z = \cases{ < \infty, & $d = 2$,\cr
< 1 + \frac{1}{d-2} \le 2, & $d \ge 3$.\cr}
\eq
This is a unexpected result, but seems to agree with some numerical solutions found in other theories of gravity \cite{Mann}.

Rescaling the coordinates $t, r, x^A$, without loss of generality, one can always set $L = \ell = 1$.
Then,  we  find that the corresponding curvature $R$ is
given by
\bqn
\lb{case1R}
R=-\frac{2d(d+1)\Lambda
(\beta-\gamma_1)^2}{\gamma_1(\beta-d\beta+d\gamma_1)(\gamma_1-d\beta+d\gamma_1)},
\eqn
which is a constant.

It is remarkable to note  that when $r_s^2>0$, $r_*=\pm r_s$ is also a solution of Eq. (\ref{hami3r}). In this case we have the same
Lifshitz solution (\ref{Lifshitz Vacuum}) but $z$ and $\Lambda$ now are given by,
\bqn
\lb{3.18a}
z&=& \frac{s(1\pm r_s)}{1-s} = - \frac{d\gamma_1}{\beta}\Bigg\{1  \pm \left[1 - \frac{(d-1)\beta}{d\gamma_1}\right]^{1/2}\Bigg\},\nb\\
\Lambda&=&0,\;\;\; \left(r_* = \pm r_s,\; r_s^2>0\right).
\eqn
%

 \subsection{Generalized BTZ Black Holes}

When $s=1$, we find that $\beta=0$. Then, from the stability conditions (\ref{2.20b}) we can see that this is possible only when $\lambda = 1$. Thus, we obtain
\bqn
\lb{case3gf}
g^2(r)&=&\frac{d(d+1)\gamma_1 }{2}\frac{r^{d+1}}{M+\Lambda r^{d+1}},\nb\\
f(r)&=&f_0r^{\frac{1-d-2z}{2}}\sqrt{M+\Lambda
r^{d+1}},
 \eqn
for which the metric takes the form,
\bqn
\lb{GBTZmetric}
\nb ds^2 &=& - N_0^2r^{1-d}\left|M \pm \left(\frac{r}{\ell}\right)^{d+1}\right|dt^2\\
\nb &&+ \frac{d(d+1)\gamma_1}{2}\left(\frac{r^{d-1}dr^2}{M \pm \left(\frac{r}{\ell}\right)^{d+1}}\right)   + r^2 d\vec{x}^2,\\
\eqn
where ``+'' (``-") corresponds to $\Lambda > 0$ ($\Lambda < 0$), and $\ell \equiv |\Lambda|^{-\frac1{d+1}}$.
Since $\gamma_1 < 0$ [cf. Eq.(\ref{2.20b})], we find that, to have $g_{rr}$ non-negative,  we must require
\bq
\lb{Gsign}
M \pm \left(\frac{r}{\ell}\right)^{d+1}  \le 0.
\eq
For $M > 0$, the above is possible only when $\Lambda < 0$, for which, by rescaling $t, r$ and $x^A$, the metric (\ref{GBTZmetric}) can be cast in the form,
\bqn
\lb{3.23a}
 ds^2 &=& L^2\left\{-\frac{\left(\frac{r}{\ell}\right)^{d+1} -M}{r^{d-1}} dt^2\right. \nb\\
 &&~~~~~~ \left. + \frac{r^{d-1}dr^2}{\left(\frac{r}{\ell}\right)^{d+1} -M}   + r^2 d\vec{x}^2\right\}, (\Lambda < 0), ~~~~~
\eqn
which is nothing but the $(d+2)$-dimensional BTZ black holes \cite{BTZ} with the black hole mass given by $M$, where $L^2 \equiv d(d+1)|\gamma_1|/2$.

It should be noted that the original  BTZ black hole  was obtained in general relativity, for which we have $(\lambda, \; \gamma_1, \beta)_{GR} = (1, -1, 0)$.
Clearly, the above solutions are valid for any given $\gamma_1 < 0$. In this sense we refer these black holes to as the generalized BTZ black holes.

Note that when $r_s^2 = 0$, we obtain $\beta = d\gamma_1/(d-1)$. Then, substituting it into the expression for $s$
we obtain $s = (d-1)/(d-2)$. However, the condition (\ref{sCondt}) require $s < (d-1)/(d-2)$. Therefore, in the current
case, $r_s$ cannot vanish.

On the other hand, when $r_s^2 < 0$, from Eqs.(\ref{sCondt}) and (\ref{pms}), we find that this is possible only when
 $ s < 0$, which is not allowed by Eq.(\ref{sCondt}). Therefore, $r_s^2 < 0$ is also impossible in the current case.
Thus, in the rest of this paper, we only need to consider the case $r_s^2 > 0$, which will be studied in the next
section.

\section{Static Spacetimes  for $r^2_s >0$ }
\renewcommand{\theequation}{4.\arabic{equation}} \setcounter{equation}{0}

The condition $r_s^2>0$ implies,
\bq
\lb{spa}
0<s<\frac{d-1}{d-2}.
\eq
However, Eq.(\ref{sCondt}) further exclude the region $0 < s < 1$.  Therefore, in this section we need only to consider the case where
\bq
\lb{spaA}
1\le s<\frac{d-1}{d-2}.
\eq
Then,   Eq.(\ref{hami3r}) can be cast in the form,
\bqn
\lb{case2hami1}
\frac{dr}{r}=\left(\frac{r_s+ {\cal
D}}{r_*+r_s}+\frac{r_s- {\cal
D}}{r_*-r_s}-\frac{2r_s}{r_*+ {\cal
D}}\right)\frac{dr_*}{2r_s{\cal P}},
\eqn
where
\bq
\lb{3.23}
{\cal P} \equiv \frac{{\cal D}^2-r_s^2}{s-1}\Delta.
\eq
Thus, from Eq.(\ref{case2hami1}) we obtain
 \bqn
  \lb{case2res1}
r(r_*)=r_H |r_*+r_s|^{\frac{r_s+ {\cal D}}{2r_s{\cal P}}}|r_*-r_s|^{\frac{r_s- {\cal D}}{2r_s{\cal P}}}|r_*+ {\cal D}|^{-\frac{1}{\cal P}},\nb\\
 \eqn
while from Eq.(\ref{case2f}) we get
\bqn
\lb{case2f1a}
\frac{df}{f}=\left(\frac{\delta_1}{r_*-r_s}+\frac{\delta_2}{r_*+r_s}+\frac{\delta_3}{r_*+ {\cal
D}}\right)dr_*,
\eqn
where
\bqn
\lb{case2f1b}
\delta_1 &\equiv&\frac{s-z+sz+  s r_s}{2r_s\Delta (r_s+ {\cal
D})},\nb\\
\delta_2 &\equiv&\frac{s-z+sz-  s r_s}{2r_s\Delta
(r_s- {\cal D})},\nb\\
\delta_3 &\equiv&\frac{z-s+s{\cal D}-zs}{\Delta
(r_s^2- {\cal D}^2)}.
\eqn
Thus,   the general solution of $f$ is given by
\bqn
\lb{case2f1c}
f=f_0|r_*-r_s|^{\delta_1}|r_*+r_s|^{\delta_2}|r_*+ {\cal D}|^{\delta_3}.
\eqn
Therefore, the metric can be rewritten in the form,
\bqn
\lb{case2metric1}
ds^2 &=& - N^2dt^2+ G^2dr_*^2 + r^2d\vec{x}^2, ~~~
\eqn
where
\bqn
\lb{case2metric2}
N^2(r_*)&=&N_0^2\left|\frac{r_*-r_s}{r_*+  {\cal D}}\right|^\frac{s(1+ r_s)}{\Sigma_+ } \left|\frac{r_*+r_s}{r_*+  {\cal D}}\right|^\frac{s(1- r_s)}{\Sigma_- },\nb\\
G^2(r_*)&=& G^2_0 \frac{(d-1)\beta+d\gamma_1(r_*^2-1)}{(r_*^2-r_s^2)^2(r_*+  {\cal D})^2},\nb\\
r^2(r_*)&=&r_H ^2\left|\frac{r_*-r_s}{r_*+  {\cal D}}\right|^\frac{1-s}{\Sigma_+} \left|\frac{r_*+r_s}{r_*+  {\cal
D}}\right|^\frac{1-s}{\Sigma_-},
\eqn
where
\bqn
\lb{3.31}
\Sigma_{\pm} &\equiv& d(1-s) + s(1 \pm r_s),\nb\\
G^2_0 &\equiv& \frac{d\gamma_1(s-1)^2[d-1+(2-d)s]^2}{2\beta\Lambda s^2(s-sd+d)^2}.
\eqn

The  corresponding Ricci scalar  is given by
\bqn
\lb{RS}
R&=&
\frac{2\beta\Lambda[2\Delta(r_*+  {\cal D})r_*-(d+1)(1-s)]}{d\gamma_1^2(1-s)(r_*^2-r_s^2)}. ~~~~
\eqn
Therefore, the spacetime is singular at $r_*=\pm r_s$.   In fact, near $r_* \simeq \pm r_s$ we find that
\bqn
\lb{r-zero}
ds^2&\simeq &\left(\frac{r}{L_{\pm}}\right)^{\frac{2s(1\pm r_s)}{1-s}}\Bigg[-d\hat{t}^2  +  \hat{G}^2_0 \left(\frac{r}{L_{\pm}}\right)^{2(d-1)}dr^2\Bigg] \nb\\
&& +r^2d\vec{x}^2,
\eqn
 where $\hat{t} \equiv \tilde{L}_{\pm} t$, and
\bqn
\epsilon^{\pm}&=& \mbox{sign} (r_*\mp r_s),\nb\\
L_{\pm}&=&r_H |2r_s|^{\frac{r_s\pm {\cal D}}{2r_s{\cal P}}}|r_s\pm {\cal D}|^{-\frac1{\cal P}},\nb\\
\tilde{L}_{\pm} &=& N_0 |2r_s|^{\frac{s(1\mp r_s)}{2\Sigma_{\mp}}}|r_s\pm {\cal D}|^{\frac{s(s-1)}{(d+s-ds)^2-s^2r_s^2}},\nb\\
\hat{G}^2_0 &=& \frac{d^2\gamma_1^2\epsilon^{\pm}r_s}{\pm \beta\Lambda L_{\pm}^2}.
\eqn
On the other hand, as $r_*\rightarrow -{\cal D}$, we have
\bq
r\rightarrow {\hat r}_0|r_*+{\cal D}|^{-\frac1{\cal P}},
\eq
where
$
{\hat r}_0=r_H |r_s-{\cal D}|^{\frac{1-s}{2\Sigma_{-}}}|r_s+{\cal D}|^{\frac{1-s}{2\Sigma_{+}}}.
$
Thus, we find that  the metric takes the asymptotical form
\bq\lb{r-infty}
ds^2\simeq -r^{2z}d{\hat t}^2+\frac{dr^2}{r^2}+r^2d\vec{x}^2,
\eq
which is precisely the Lifshitz space-time (\ref{Lifshitz Vacuum}) with
\bqn
\lb{r-infty.A}
z &=& \frac{s}{s(1-d) +d},\nb\\
\hat{t} &=& N_0 {\hat r}_0^{-\frac{s}{s-ds+d}} |r_s+{\cal D}|^{\frac{s(1+r_s)}{2\Sigma_{+}}}|r_s-{\cal D}|^{\frac{s(1-r_s)}{2\Sigma_{-}}} t. ~~~~
\eqn
Note that in writing the above metric we had  used the
condition
\bq
\lb{asmptS}
d^2\gamma_1^2({\cal D}^2 -r_s^2) = 2\beta\Lambda.
\eq

To study the above solutions further, let us consider the cases with different values of $s$ separately.

\subsection{$1<s<\frac{d}{d-1}$}

In this case, we have
 \bqn
\lb{scase3}
r(r_*) = \cases{r_H , & $r_* \rightarrow -\infty$,\cr
\infty, & $r_*  = - {\cal D}$,\cr
0, & $r_*  = -  r_s$,\cr
\infty, & $r_*  = +  r_s$,\cr
r_H , & $r_* \rightarrow + \infty$.\cr}
 \eqn

  \begin{figure}[tbp]
\centering
\includegraphics[width=8cm]{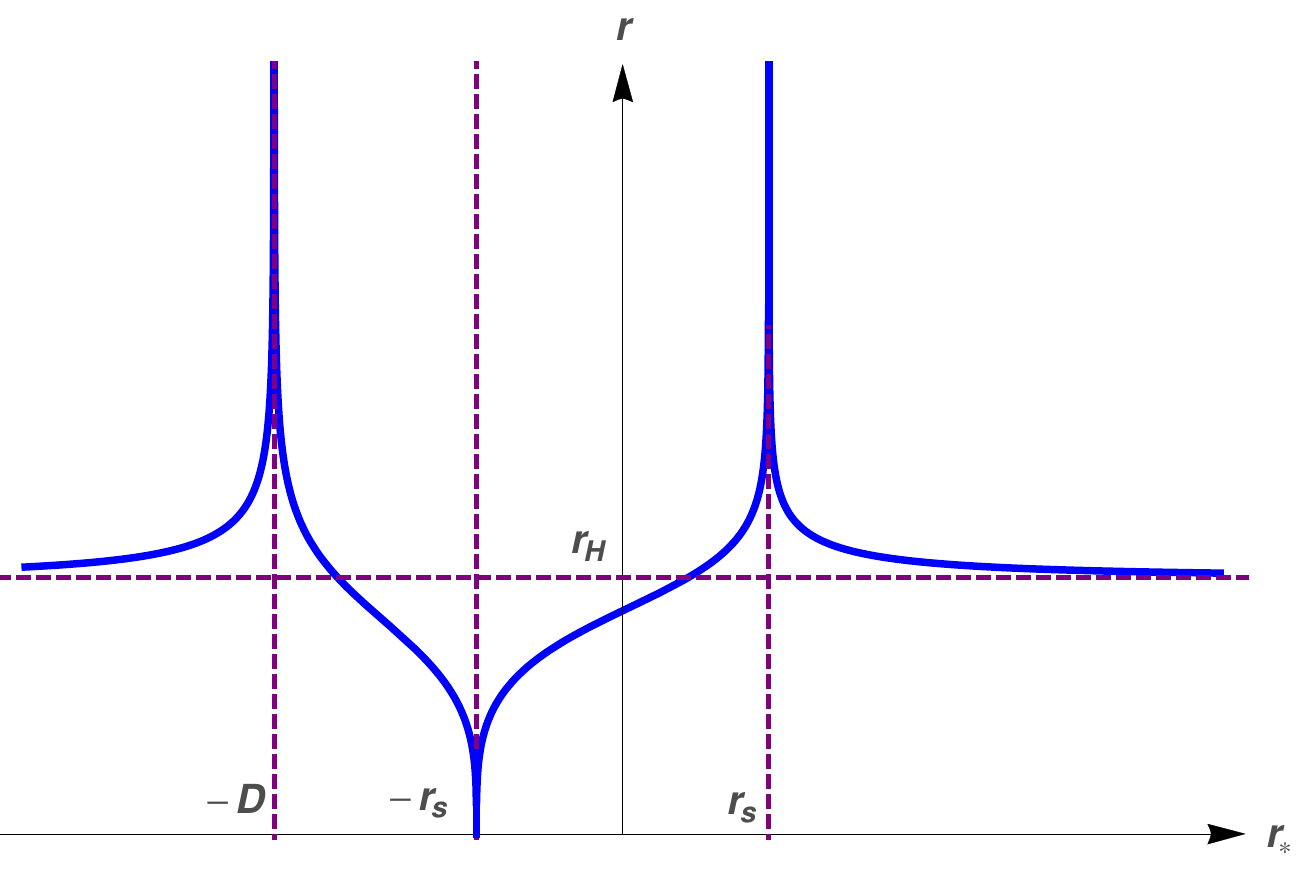}
\caption{The function $r\equiv r (r_*)$ for $r_s^2 > 0$ and $1<s<\frac{d}{d-1}$, where $D \equiv {\cal{D}}$. The spacetime is singular at $r_*=\pm r_s$, and
 asymptotically Lifshitz as $r_* \rightarrow - {\cal{D}}$.}
\label{fig1}
\end{figure}

Fig. \ref{fig1} shows the function $r(r_*)$ vs $r_*$, from which we can see that the region $r \in [0, \infty)$ is mapped into the region $r_* \in [-r_s, +r_s)$ or
$r_* \in (-{\cal D},  -r_s]$. The region $r_* \in (-\infty, -{\cal D})$ or $r_* \in (r_s, +\infty)$ is mapped into the one $r \in (r_H , +\infty)$.

As shown before, the space-time is singular at $r_* = \pm r_s$, and as $r \rightarrow \infty$
(or $r_* \rightarrow -{\cal D}$), it is asymptotically approaching to the Lifshitz space-time (\ref{Lifshitz Vacuum}) with $z = s(d+s-sd)^{-1}$.

To study the solutions further, let us rewrite Eq. (\ref{case2res1}) in the form
\bqn
\lb{Rstar}
\left(\frac{r}{r_H }\right)^{\hat s} &=& \frac{({\cal D}-r_s)\epsilon^-}{{\cal D}+r_s}
\left(\epsilon^+\mathfrak{R}^{\frac{2r_s}{r_s-{\cal D}}}+\frac{2\epsilon^{\cal D}r_s}{{\cal D}-r_s}\mathfrak{R}\right), ~~~~~~~~
\eqn
where $\epsilon^{\cal D}\equiv {\mbox{sign}}(r_*+{\cal D})$ and
\bqn
\lb{Rstar_b}
\mathfrak{R}\equiv \left|\frac{r_* -r_s}{r_*+ {\cal D}}\right|^{\frac{r_s-{\cal D}}{r_s+{\cal D}}},\;\;\;
{\hat s}\equiv \frac{2r_s{\cal P}}{r_s+{\cal D}}.
\eqn
It should be noted that the above two equations are valid for any $1\le s <\frac{d-1}{d-2}$.
As a representative example, let us consider the case
${\cal D}=3r_s$, which corresponds to
\bq
\lb{sp3}
s=\frac{-1-17d+18d^2-\sqrt{1+34d+d^2}}{2(7-17d+9d^2)}.
\eq
Thus, Eqs.(\ref{Rstar}) and (\ref{Rstar_b}) reduce to,
\bqn
\lb{3.48}
\left(\frac{r}{r_H }\right)^{\hat s} &=& \frac{\epsilon^-}{2\mathfrak{R}}\left(\epsilon^+ +\epsilon^{\cal D}\mathfrak{R}^2\right), \nb\\
\mathfrak{R} &=&  \left|\frac{\tilde{r}_* +3}{\tilde{r}_*-1}\right|^{1/2}.
\eqn

To study the solutions further, let us consider the following
cases, separately.

(a) ${r}_*\in (-\infty, -{\cal{D}}]$,   we have  $\epsilon^+=\epsilon^-=\epsilon^{\cal D}=-1$. Then, from Eq.(\ref{3.48}) we obtain
\bqn
\mathfrak{R}&=&\left(\frac{r}{r_H }\right)^{\hat s}\left(1\pm\sqrt{1-\left(\frac{r_H }r\right)^{2\hat s}}\right),\nb\\
r_*&=& \frac{\mathfrak{R}^{2} + 3}{\mathfrak{R}^{2} - 1}.
\eqn
Since $\mathfrak{R} \in [0, 1)$, as it can be seen from Eq.(\ref{3.48}), we find that only the root $\mathfrak{R}_-$ satisfies this condition.
On the other hand, from Eqs.(\ref{case2f1c}) and (\ref{case2g}) we find,
\bqn
r^{2z}f^2&=& \frac{N_0^2}{ \mathfrak{R}^3_-}\left(\frac{r}{r_H }\right)^{\frac{3{\hat s}(r_s-1)}{2r_s}},\\
g^2 &=&  \frac{1+ \mathfrak{R}_-^2}{\left(\mathfrak{R}_-^2 -1\right)^{2}},
\eqn
where
\bq
\lb{3.51a}
\mathfrak{R}_- = \frac{\left(\frac{r_H }{r}\right)^2}{1 + \sqrt{1 -\left(\frac{r_H }{r}\right)^4}}
= \cases{1, & $r = r_H $,\cr
0, & $ r = \infty$.\cr}
\eq

(b)  ${r}_*\in (-{\cal{D}},-r_s]$, we have $\epsilon^+=\epsilon^-=-\epsilon^{\cal D}=-1$, and now $\mathfrak{R}\in (0,1]$. Thus we find that
\bqn
\mathfrak{R}&=&\left(\frac{r}{r_H }\right)^{\hat s}\left(\sqrt{1+\left(\frac{r_H }r\right)^{2\hat s}}-1\right),\nb\\
\tilde{r}_*&=& \frac{\mathfrak{R}^{2} - 3}{\mathfrak{R}^{2} + 1},
\eqn
are  solutions to Eq. (\ref{Rstar}) in this region. This immediately  leads to,
\bqn
r^{2z}f^2&=& \frac{N_0^2}{ \mathfrak{R}^3}\left(\frac{r}{r_H }\right)^{\frac{3{\hat s}(r_s-1)}{2r_s}},\\
g^2 &=&  \frac{1- \mathfrak{R}^2}{\left(\mathfrak{R}^2 +1\right)^{2}}.
\eqn

(c) ${r}_*\in (-r_s, r_s]$, we have $-\epsilon^+=\epsilon^-=\epsilon^{\cal D}=1$, implying $\mathfrak{R}\in (1,+\infty)$. Then,  we find that
\bqn
\mathfrak{R}&=&\left(\frac{r}{r_H }\right)^{\hat s}\left(\sqrt{1+\left(\frac{r_H }r\right)^{2\hat s}}+1\right),\nb\\
\eqn
are  solutions to Eq. (\ref{Rstar}). We therefore obtain the same forms as region (b) for functions $g$ and $f$
\bqn
r^{2z}f^2&=& \frac{N_0^2}{ \mathfrak{R}^3}\left(\frac{r}{r_H }\right)^{\frac{3{\hat s}(r_s-1)}{2r_s}},\\
g^2 &=&  \frac{1- \mathfrak{R}^2}{\left(\mathfrak{R}^2 +1\right)^{2}}.
\eqn

(d) ${r}_*\in [r_s,+\infty)$, we have $\epsilon^+=\epsilon^-=\epsilon^{\cal D}=1$, implying $\mathfrak{R}\in (1,+\infty)$. Then,  we find that
\bqn
\mathfrak{R}&=&\left(\frac{r}{r_H }\right)^{\hat s}\left(\sqrt{1-\left(\frac{r_H }r\right)^{2\hat s}}+1\right),\nb\\
\eqn
are  solutions to Eq. (\ref{Rstar}). We therefore obtain the functions $g$ and $f$
\bqn
r^{2z}f^2&=& \frac{N_0^2}{ \mathfrak{R}^3}\left(\frac{r}{r_H }\right)^{\frac{3{\hat s}(r_s-1)}{2r_s}},\\
g^2 &=&  \frac{1+ \mathfrak{R}^2}{\left(\mathfrak{R}^2 -1\right)^{2}}.
\eqn

 \subsection{ $s=\frac{d}{d-1}$}

In this case,  we find that $\beta=\gamma_1$. Then, we obtain
\bqn
\lb{casenew1fg}
g^2(r)&=&\frac{2d\gamma_1g_0r^{2 \sqrt{d}}}{\Lambda \left(r^{2 \sqrt{d}} - g_0\right)^2},\nb\\
 f(r)&=&\frac{f_0 r^{-d-z+ \sqrt{d}}}{r^{2 \sqrt{d}}-g_0},
\eqn
where $f_0$ and $g_0$ are two integration constants. Then, the corresponding metric takes the form
\bqn
\lb{casenew1fg.A}
ds^2 &=& - f_0^2\frac{r^{2(\sqrt{d} - d)}dt^2}{\left(r^{2\sqrt{d}} - g_0\right)^2} + \left(\frac{2d\gamma_1 g_0}{\Lambda}\right)\frac{r^{2(\sqrt{d} -1)}dr^2}{\left(r^{2\sqrt{d}} - g_0\right)^2}\nb\\
&&
+ r^2d\vec{x}^2.
\eqn
Clearly, to have $g_{rr}$ positive, we must assume that
\bq
\lb{3.27}
\frac{\gamma_1g_0}{\Lambda} > 0.
\eq
The corresponding Ricci scalar  is given by
\bqn
\lb{3.28}
R&=&\frac{\Lambda r^{-2\sqrt{d}}}{2g_0\gamma_1}\left(r^{2\sqrt{d}}-g_0\right)\left[(1-\sqrt{d})^2g_0\right.\nb\\
&& ~~~~~~~~~~~~~ \left. -(1+\sqrt{d})^2r^{2\sqrt{d}}\right].
\eqn
which remains finite at  the hypersurface $r= r_H$, and  indicate that it might represent a horizon, where $r_H = g_0^{1/(2\sqrt{d})}$.
As $r \rightarrow \infty$, the metric takes the following
asymptotical form,
\bqn
\lb{3.29}
ds^2 &\simeq& \left(\frac{\tilde{r}_0}{ \tilde{r}}\right)^{\frac{2}{1+\sqrt{d}}}
\left[-\left(\frac{\tilde{r}}{\tilde{r}_0}\right)^{\frac{2(\sqrt{d}+d+1)}{1+\sqrt{d}}}d\tilde{t}^2 + d\tilde{r}^2 + d\vec{x}^2\right],\nb\\
\eqn
where $\tilde{t} = f_0 t$, and
\bq
\lb{3.30}
\tilde{r} =  \tilde{r}_0 r^{-(1+\sqrt{d})},\;\;\;  \tilde{r}_0 \equiv \frac{\sqrt{2d\gamma_1 g_0/\Lambda}}{\sqrt{d} +1}.
\eq
Rescaling $\tilde{t}, \tilde{r}$ and $x^{A}$, the above metric can be cast in the form,
\bq
\lb{3.31A}
ds^2 \simeq \hat{r}^{-\frac{2(d-\theta)}{d}}\left(- \hat{r}^{-2(z-1)}d\hat{t}^2 + d\hat{r}^2 + d\hat{\vec{x}}^2\right),
\eq
where
\bq
\lb{3.32}
\theta = \frac{d\sqrt{d}}{1+\sqrt{d}},\;\;\; z = - \frac{d}{1+\sqrt{d}}.
\eq
The metric (\ref{3.31A}) is nothing but the space-time with non-relativistic scaling and hyperscaling violation. It was first constructed in Einstien-Maxwell-dilaton theories \cite{GR10},
and recently has been extensively studied in \cite{OTU}. Under the anisotropic scaling (\ref{1.1}), it is not invariant but rather scaling as $ds^2 \rightarrow b^{2\theta/d}ds^2$.
This kind of non-relativistic  scaling is closely related to the existence of Fermi surfaces, in which the entanglement entropy is logarithmically proportional to the erea, $S \simeq A \log A$.

\subsection{$\frac{d}{d-1}<s<\frac{d^2}{d^2-d-1}$}

\begin{figure}[tbp]
\centering
\includegraphics[width=8cm]{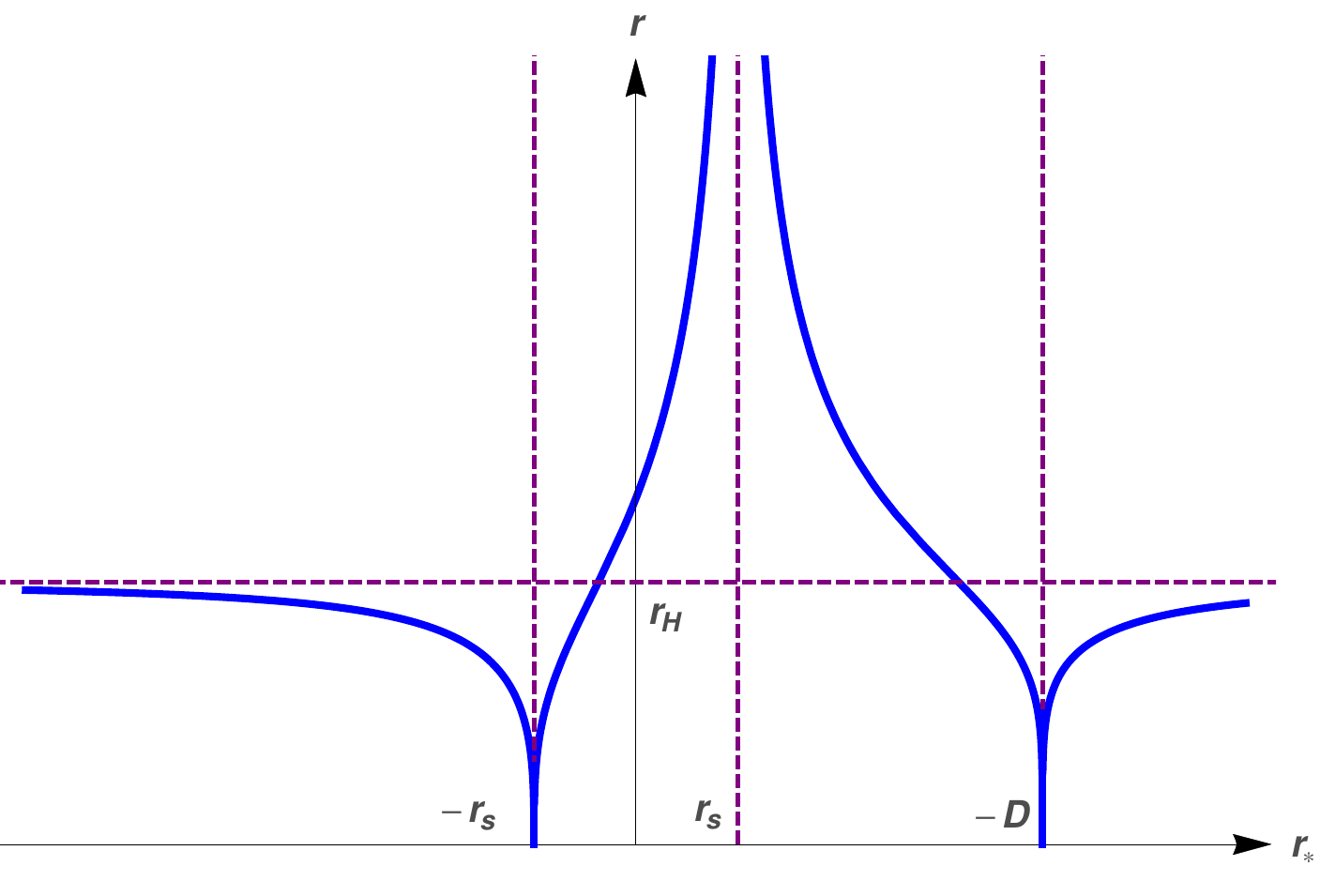}
\caption{The function $r\equiv r (r_*)$ for $r_s^2 > 0$ and
$\frac{d}{d-1}<s<\frac{d^2}{d^2-d-1}$, where $D \equiv {\cal{D}} < -r_s$ in the present case. The spacetime is singular at $r_* = \pm r_s$.}
\label{fig2}
\end{figure}

In this case, we have
 \bqn
\lb{scase4}
r(r_*) = \cases{r_H , & $r_* \rightarrow -\infty$,\cr
0, & $r_*  = -  r_s$,\cr
\infty, & $r_*  = +  r_s$,\cr
0, & $r_*  = - {\cal D}$,\cr
r_H , & $r_* \rightarrow + \infty$.\cr}
 \eqn
Note that in the current case we have  ${\cal{D}} < - r_s < 0$.
Fig. \ref{fig2} shows the function $r(r_*)$ vs $r_*$, from which we can see that the region $r \in [0, \infty)$ is mapped into the region $r_* \in [-r_s, +r_s)$ or
$r_* \in (r_s,-{\cal D}]$. The region $r_* \in (-\infty, -r_s)$ or $r_* \in (-{\cal D}, +\infty)$ is mapped into the one $r \in (r_H , +\infty)$.

Similar to the previous cases, let us consider the  case with ${\cal D}=-3r_s$ in detail, which corresponds to
\bq
\lb{sp4}
s=\frac{-1-17d+18d^2+\sqrt{1+34d+d^2}}{2(7-17d+9d^2)}.
\eq
 Then,  we find that
\bqn
\lb{3.76}
&& \left(\frac{r}{r_H }\right)^{\hat s} =2\epsilon^-\left(\epsilon^+ \mathfrak{R}^{\frac12}-\frac{\epsilon^{\cal D}}{2}\mathfrak{R}\right),\nb\\
&& \mathfrak{R} = \left(\frac{\tilde{r}_* -3}{\tilde{r}_* - 1}\right)^2.
\eqn
Following what we did for the previous cases, one can solve it for $\mathfrak{R}$   in the following four regions.

(a)  ${r}_*\in(-\infty, -r_s]$. In this region, we have the following solution
\bqn
\nb\mathfrak{R}^{\frac12}&=&1+\sqrt{1-\left(\frac{r}{r_H }\right)^{\hat s}}.
\eqn
Then, the functions $f$ and $g$ are given by
\bqn
f^2&=&N_0^2r^{-2z}\mathfrak{R}^{-\frac32}\left(\frac{r}{r_H }\right)^{\frac{3(r_s-1){\hat s}}{4r_s}},\\
g^2& =&  \frac{2-\mathfrak{R}^{\frac12}}{2\left(1-\mathfrak{R}^{\frac12}\right)^2},
\eqn

(b)  ${r}_*\in(-r_s, r_s]$. In this region, we have the following solution
\bqn
\nb\mathfrak{R}^{\frac12}&=&1+\sqrt{1+\left(\frac{r}{r_H }\right)^{\hat s}}.
\eqn
Then, the functions $f$ and $g$ are given by
\bqn
f^2&=&N_0^2r^{-2z}\mathfrak{R}^{-\frac32}\left(\frac{r}{r_H }\right)^{\frac{3(r_s-1){\hat s}}{4r_s}},\\
g^2& =&  \frac{2-\mathfrak{R}^{\frac12}}{2\left(1-\mathfrak{R}^{\frac12}\right)^2},
\eqn

(c)  ${r}_*\in(r_s, {\cal{D}}]$. In this region, we have the following solution
\bqn
\nb\mathfrak{R}^{\frac12}&=&-1+\sqrt{1+\left(\frac{r}{r_H }\right)^{\hat s}}.
\eqn
Then, the functions $f$ and $g$ are given by
\bqn
f^2&=&N_0^2r^{-2z}\mathfrak{R}^{-\frac32}\left(\frac{r}{r_H }\right)^{\frac{3(r_s-1){\hat s}}{4r_s}},\\
g^2& =&  \frac{2+\mathfrak{R}^{\frac12}}{2\left(1+\mathfrak{R}^{\frac12}\right)^2},
\eqn

(d)  ${r}_*\in[{\cal{D}},+\infty)$. In this region, we have the following solution
\bqn
\nb\mathfrak{R}^{\frac12}&=&1-\sqrt{1-\left(\frac{r}{r_H }\right)^{\hat s}}.
\eqn
Then, the functions $f$ and $g$ are given by
\bqn
f^2&=&N_0^2r^{-2z}\mathfrak{R}^{-\frac32}\left(\frac{r}{r_H }\right)^{\frac{3(r_s-1){\hat s}}{4r_s}},\\
g^2& =&  \frac{2-\mathfrak{R}^{\frac12}}{2\left(1-\mathfrak{R}^{\frac12}\right)^2},
\eqn

 \subsection{$s=\frac{d^2}{d^2-d-1}$}

\begin{figure}[tbp]
\centering
\includegraphics[width=8cm]{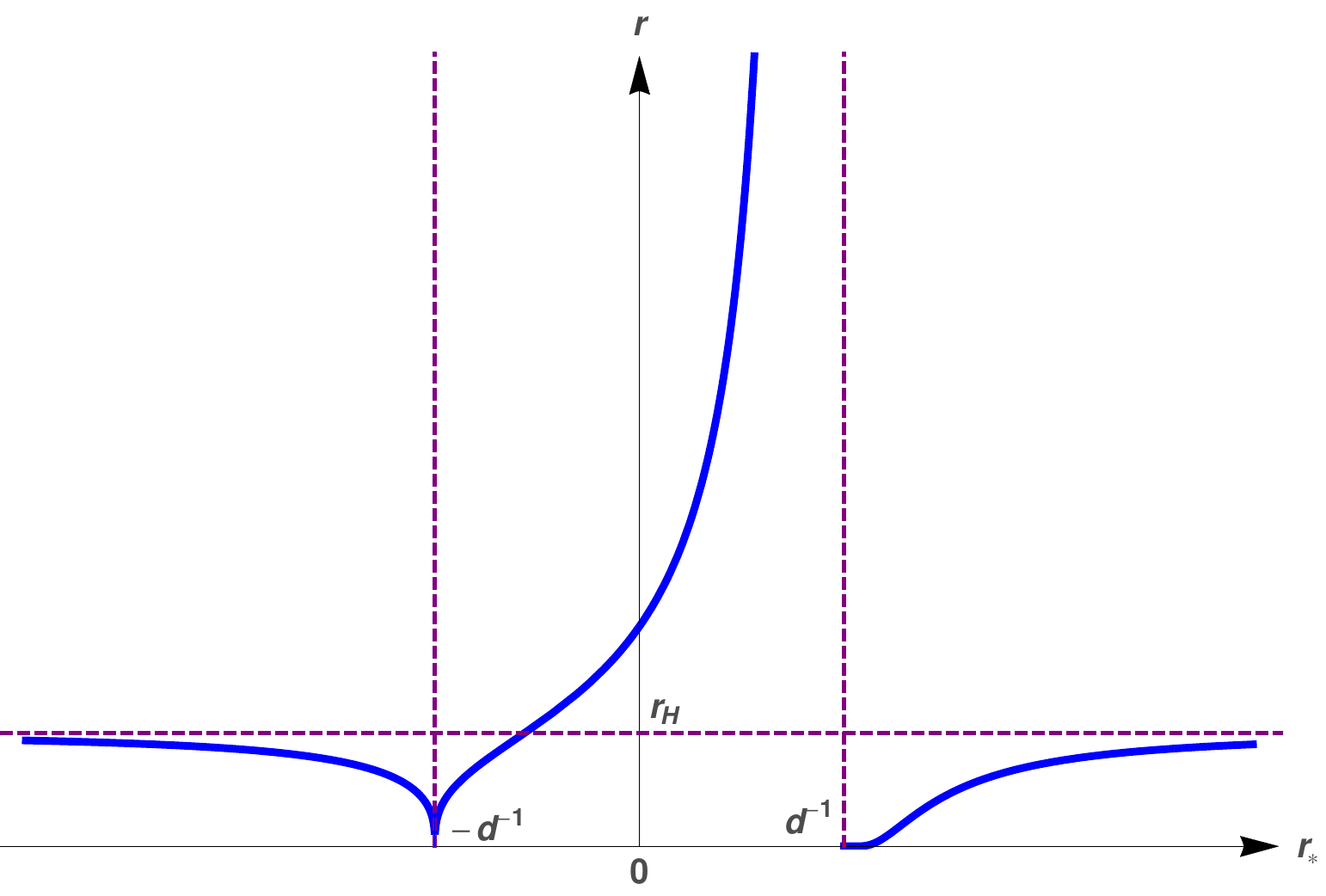}
\caption{The function $r\equiv r (r_*)$ for $s=\frac{d^2}{d^2-d-1}$.
The space-time is singular at $r_* = \pm d^{-1}$, as can be seen
from Eq.(\ref{casenew2RicciaA}).}
 \label{fig3}
\end{figure}

When
\bq
\lb{3.33}
s=\frac{d^2}{d^2-d-1},
\eq
we find that $\beta=\gamma_1\frac{d+1}{d}$, and
Eq.(\ref{hami3r}) becomes
\bqn
 \lb{casenew2Hami}
r'_*=\frac{d^3(r_*-d^{-1} )(r_*^2-d^{-2})}{(d+1)r}.
\eqn
To solve the above equation, we first write the above equation in the form,
\bqn
 \lb{casenew2Hamia}
\frac{dr}{r}&=& \frac{d+1}{2d^2}\left[\frac{1}{(r_*-d^{-2})^2}-\frac{d/2}{r_*-d^{-1}}\right.\nb\\
&& ~~~ \left. +\frac{d/2}{r_*+d^{-1}}\right]dr_*,
\eqn
which has the general  solution,
\bqn
\lb{casenew2ra}
r=r_H \left|\frac{r_*+ d^{-1}}{r_*-d^{-1}}\right|^{\frac{d+1}{4d}}e^{-\frac{d+1}{2d^2(r_*-d^{-1})}}.
\eqn
Thus, we have
\bqn
\lb{scase7}
r(r_*) = \cases{r_H , & $r_* \rightarrow -\infty$,\cr
\infty, & $\left(r_*- d^{-1}\right)  \rightarrow 0^-$,\cr
0, & $\left(r_*- d^{-1}\right)  \rightarrow 0^+$,\cr
0, & $r_*  = - d^{-1}$,\cr
r_H , & $r_* \rightarrow + \infty$.\cr}
 \eqn
Fig. \ref{fig3} shows the curve of $r$ vs $r_*$. From the definition
of $W(r)$, on the other hand, we find that
 \bqn
\lb{casenew2Fa}
\frac{df}{f}&=&\bigg[-\frac{d^2+d+z+dz}{2(dr_*-1)^2}+\frac{d^2-d+z+dz}{4(dr_*-1)}\nb\\
&&-\frac{d^2-d+z+dz}{4(dr_*+1)}\bigg]dr_*,
\eqn
which has the general solution,
\bq
\lb{3.39}
f =f_0\left|\frac{r_*-d^{-1}}{r_*+d^{-1}}\right|^{\frac{d^2-d+z+dz}{4d}}\exp{\left[\frac{(d+1)(d+z)}{2d^2(r_*-d^{-1})}\right]}.
\eq
Therefore, the corresponding metric  takes the form,
\bqn
\lb{casenew2metricA}
ds^2= -N^2(r_*) dt^2+G^2(r_*) dr_*^2+r^2(r_*) d\vec{x}^2,
\eqn
where
 \bqn
  \lb{casenew2metrica}
N^2&=&N_0^2\left|\frac{r_*-\frac{1}{d}}{r_*+\frac{1}{d}}\right|^{\frac{d-1}{2}}\exp{\left[\frac{d+1}{d(r_*-d^{-1})}\right]},\nb\\
G^2&=&\frac{(d+1)\gamma_1}{2d^3 \Lambda (r_*-d^{-1})^3(r_*+d^{-1})},
 \eqn
 where $r(r_*)$ is given by Eq.(\ref{casenew2ra}). Then, the corresponding Ricci scalar is given by
 \bqn
  \lb{casenew2RicciaA}
R&=& \frac{4\Lambda d}{\gamma_1(r_*^2-d^{-2})}\left[r_*(r_*-d^{-1}) + \frac{(d+1)^2}{2d^3}\right], ~~~~~
 \eqn
from which it can be seen that the space-time is singular at $r_* = \pm d^{-1}$. Then, the physical interpretation of the solutions
in the region $- d^{-1} \le r_* \le d^{-1}$ is not clear. On the other hand, to have a complete space-time in  $ r_* \in (-\infty, -d^{-1})$
or  $ r_* \in (d^{-1}, \infty)$, extensions beyond the hypersurfaces $r_* = \pm \infty$ are needed.

\subsection{$\frac{d^2}{d^2-d-1}<s<\frac{d-1}{d-2}$}

\begin{figure}[tbp]
\centering
\includegraphics[width=8cm]{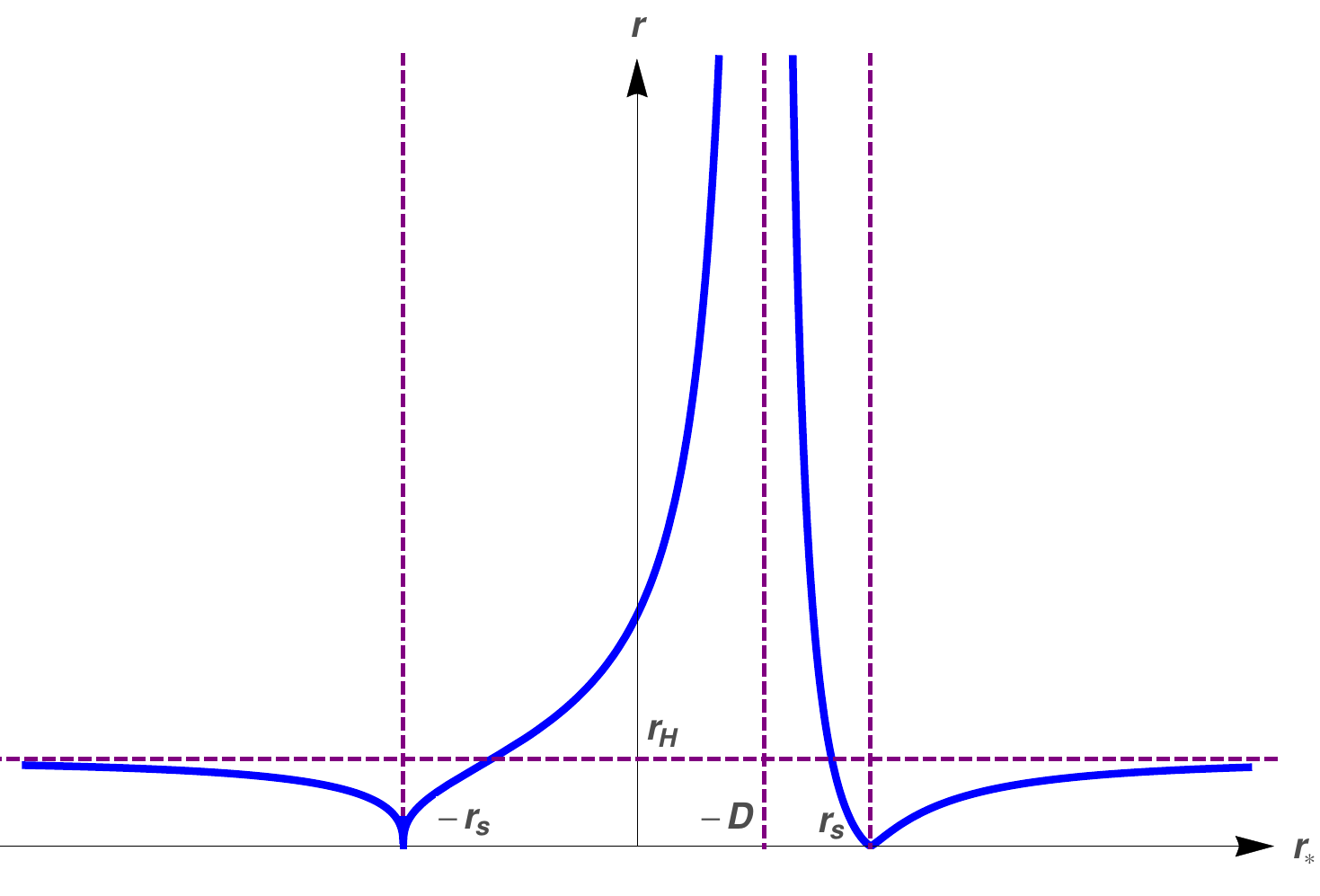}
\caption{The function $r\equiv r (r_*)$ for
$\frac{d^2}{d^2-d-1}<s<\frac{d-1}{d-2}$, where now $-r_s < D \equiv {\cal{D}} < 0$. The spacetime is singular at $r_* = \pm r_s$,
and  asymptotically Lifshitz as $r_* \rightarrow - {\cal{D}}$.}
\label{fig4}
\end{figure}

In this case, we have
 \bqn
\lb{scase5}
r(r_*) = \cases{r_H , & $r_* \rightarrow -\infty$,\cr
0, & $r_*  = -  r_s$,\cr
\infty, & $r_*  = - {\cal D}$,\cr
0, & $r_*  = +  r_s$,\cr
r_H , & $r_* \rightarrow + \infty$.\cr}
\eqn
Similar to the last case, now   ${\cal{D}} < 0$ but with ${\cal{D}} > -r_s$.
 Fig. \ref{fig4} shows the function $r(r_*)$ vs $r_*$, from which we can see that the region $r \in [0, \infty)$ is mapped into the region $r_* \in [-r_s, -{\cal D})$ or
$r_* \in (-{\cal D},r_s]$. The region $r_* \in (-\infty, -r_s)$ or $r_* \in (r_s, +\infty)$ is mapped into the one $r \in (r_H , +\infty)$.

Similar to the previous cases, let us consider the  case with $r_s=-3{\cal D}$ in detail, which corresponds to
\bq
\lb{sp5}
s=\frac{-9+7d+2d^2+3\sqrt{9-14d+9d^2}}{2(-17+7d+d^2)}.
\eq
 Then,  we find that
\bqn
&& \left(\frac{r}{r_H }\right)^{\hat s} =-2\epsilon^-\left(\epsilon^+ \mathfrak{R}^{\frac32}-\frac{3\epsilon^{\cal D}}{2}\mathfrak{R}\right),\nb\\
&& \mathfrak{R} = \left(\frac{\tilde{r}_* -1}{\tilde{r}_* - \frac13}\right)^2.
\eqn
Following what we did for the previous cases, one can solve it for $\mathfrak{R}$   in the following four regions.

(a)  ${r}_*\in(-\infty, -r_s]$. In this region, we have the following solution
\bq
\mathfrak{R}=\frac12+\cos{\frac{2\tilde\theta}{3}} = \cases{\frac{3}{2}, & $r = r_H $,\cr
1, & $r = 0$.\cr}
\eq
where $\tilde\theta$ is defined as
\bq\lb{t-theta}
\cos{\tilde\theta}=\left(\frac{r}{r_H }\right)^{\frac{\hat s}2},\ \ \ \sin{\tilde\theta}=\sqrt{1-\left(\frac{r}{r_H }\right)^{\hat s}}.
\eq
Since $\tilde\theta \in [0, \pi/2$], we have $\mathfrak{R} \ge 1$ for $r \in [0, r_H ]$. The functions $f$ and $g$ are also given by

\bqn\lb{fg2}
f^2&=&N_0^2r^{-2z}\mathfrak{R}^{-\frac12}\left(\frac{r}{r_H }\right)^{\frac{(r_s-1){\hat s}}{4r_s}},\\
g^2& =&  \frac{2\mathfrak{R}-3\mathfrak{R}^{\frac12}}{2\left(1-\mathfrak{R}^{\frac12}\right)^2},
\eqn
from which we can see that $g$ becomes unbounded at $r = 0$  (or $\tilde{r}_* = \pm 1$).  As shown above, this is a coordinate singularity.

To extend the above solution to the region $r > r_H $, one may simply assume that Eq.(\ref{t-theta}) hold also for $r > r_H $. In particular, setting
$\tilde\theta = i \hat{\theta}$, we find that
\bq
\lb{3.85}
\mathfrak{R}=\frac12+\cosh{\frac{2\hat{\theta}}{3}} \ge \frac{3}{2},   \; (r \ge r_H ),
\eq
where $\hat{\theta}$ is defined by
\bq
\cosh{\hat{\theta}}=\left(\frac{r}{r_H }\right)^{\frac{\hat s}2},\ \ \ \sinh{\hat{\theta}}=\sqrt{\left(\frac{r}{r_H }\right)^{\hat s} - 1}.
\eq
 The above expression represents an extension of the solution originally defined only for $r \le r_H $. Note that
$\mathfrak{R} \simeq r^{4/3}$ as $r \rightarrow \infty$. Then, from Eq.(\ref{fg2}) we find that
\bqn
\lb{3.86}
&& r^{2z}f^2 \sim r^{\frac{(r_s-1){\hat s}}{4r_s}-\frac23},\;\;\;\; g^2 \simeq 1,
\eqn
as $r  \rightarrow \infty$. That is, the space-time is asymptotically approaching to a Lifshitz space-time with its dynamical exponent now given by
$$z =  \frac{(r_s-1){\hat s}}{8r_s}-\frac13.$$

(b)  ${r}_*\in(-r_s,{\cal{D}}]$. In this region, we have the following solution
\bqn
\nb\mathfrak{R}^{\frac12}&=&-\frac12+\frac12\left[\frac{r}{r_H }+\sqrt{1+\left(\frac{r}{r_H }\right)^{\hat s}}\right]^{-\frac23}\\
&&+\frac12\left[\frac{r}{r_H }+\sqrt{1+\left(\frac{r}{r_H }\right)^{\hat
s}}\right]^{\frac23}. \eqn Then, the functions $f$ and $g$ are given
by \bqn
f^2&=&N_0^2r^{-2z}\mathfrak{R}^{-\frac12}\left(\frac{r}{r_H }\right)^{\frac{(r_s-1){\hat s}}{4r_s}},\\
g^2 &=&  \frac{2\mathfrak{R}-3\mathfrak{R}^{\frac12}}{2\left(1-\mathfrak{R}^{\frac12}\right)^2}.
\eqn

(c)  ${r}_*\in({\cal{D}}, r_s]$. In this region, we have the following solution
\bqn
\mathfrak{R}^{\frac12}=\cases{-\frac12+\frac12\mathcal{A}(r)^{-\frac23}+\frac12\mathcal{A}(r)^{\frac23}, & $r\geq r_H $,\cr
-\frac12+\cos{\frac{2\tilde\theta}{3}}, & $r< r_H $,\cr}
\eqn
where we have defined
\bqn
\lb{theta2}\mathcal{A}(r)&=& \left(\frac{r}{r_H }\right)^{\frac{\hat s}{2}} +\sqrt{\left(\frac{r}{r_H }\right)^{\hat s} - 1},
\eqn
and $\tilde\theta$ is given by (\ref{t-theta}).

The functions $f$ and $g$ are given by
\bqn
f^2&=&N_0^2r^{-2z}\mathfrak{R}^{-\frac12}\left(\frac{r}{r_H }\right)^{\frac{(r_s-1){\hat s}}{4r_s}},\\
g^2 &=& \frac{2\mathfrak{R}+5\mathfrak{R}^{\frac12}}{2\left(1+\mathfrak{R}^{\frac12}\right)^2}.
\eqn

(d)  ${r}_*\in(r_s,+\infty)$. In this region, we have the following solution
\bqn
\mathfrak{R}&=& \frac12+\cos{\frac{2\tilde\theta+\pi}{3}}
= \cases{1, & $r = r_H $,\cr
0, & $r = 0$,\cr}
\eqn
where $\tilde\theta$ is defined by Eq.(\ref{t-theta}), so that $\mathfrak{R} \in (0, 1)$. Then, the  functions $f$ and $g$ are given by
\bqn
f^2&=&N_0^2r^{-2z}\mathfrak{R}^{-\frac12}\left(\frac{r}{r_H }\right)^{\frac{(r_s-1){\hat s}}{4r_s}},\\
g^2 &=&  \frac{2\mathfrak{R}-3\mathfrak{R}^{\frac12}}{2\left(1-\mathfrak{R}^{\frac12}\right)^2}.
\eqn
Clearly, the metric becomes singular at $r = r_H $.  But this singularity is just a coordinate singularity   and extension beyond this surface is needed.
Simply assuming that Eq.(\ref{t-theta}) holds also for $r > r_H $ will lead to $\mathfrak{R}$ to be a complex function of $r$, and so are
the functions $f$ and $g$. Therefore, this will not represent a desirable extension.

\section{Universal Horizons and Black Holes}
\renewcommand{\theequation}{5.\arabic{equation}} \setcounter{equation}{0}

Remarkably, studying the behavior of a khronon field in the fixed Schwarzschild black hole background,
\bq
\lb{eq1.1}
ds^2 = - \left(1- \frac{r_s}{r}\right)dv^2 + 2dvdr + r^2d\Omega^2,
\eq
where $r_s \equiv 2M$,  Blas and Sibiryakov showed that a universal horizon exists
inside the Killing horizon.      
But, in contrast to it, now the universal horizon is spacelike, and on which   
the time-translation  Killing vector $\zeta^{\mu} [ =   \delta^{\mu}_{v}]$   becomes orthogonal to    $u^{\mu}$,
\bq
\lb{eq1.2}
u_{\mu} \zeta^{\mu} = 0,
\eq
where $u_{\mu}$ is the normal unit vector of the timelike foliations $\phi\left(x^{\mu}\right) = $ Constant,  
 \bq
\lb{eq1.3}
u_{\mu} = \frac{\phi_{,\mu}}{\sqrt{X}},
\eq
with $ X \equiv -g^{\alpha\beta}\partial_{\alpha} \phi \partial_{\beta} \phi$. Since $u_{\mu}$ is well-defined in the whole space-time, and remains timelike 
from the asymptotical infinity ($r = \infty$) all the way down to the space-time singularity ($r = 0$),
Eq.(\ref{eq1.2}) is possible only inside the Killing horizon, as only there  $\zeta^{\mu}$ becomes spacelike and can be
possibly orthogonal to $u_{\mu}$.

The above definition of the universal horizons can be easily generalized to any theory that breaks Lorentz symmetry either in
the level of the action, such as  the HL gravity studied in this paper, or spontaneously, such as
the khrononmetric theory \cite{BS11},  ghost condensation  \cite{GC},  Einstein-aether theory \cite{EA} \footnote{When the aether field $u_{\mu}$ is 
hypersurface-orthoginal, $u_{[\nu}D_{\alpha}u_{\beta]} = 0$, 
where $D_{\mu}$  denotes the covariant derivative with respect to the bulk metric $g_{\mu\nu}$, the
the Einstein-aether theory is  equivalent to  the khrononmetric theory, as 
shown explicitly in  \cite{Jacobson10} [See also \cite{Wang13,BS11}].}, and massive gravity \cite{mGR}.
The idea is simply to consider  the khronon field as a probe field that plays the same role as a Killing vector field for any given space-time \cite{LACW,LGSW}.

The equation that the khronon must satisfy in a given background $g_{\mu\nu}$ can be obtained form the action \cite{LACW,LGSW},
  \bqn
 \lb{U2}
S_{\phi} &=&  \int d^{D+1}x \sqrt{|g|}\Big[c_1\left(D_{\mu}u_{\nu}\right)^2 + c_2 \left(D_{\mu}u^{\mu}\right)^2\nb\\
&& ~~  + c_3   \left(D^{\mu}u^{\nu}\right)\left(
D_{\nu}u_{\mu}\right)   - c_4 a^{\mu}a_{\mu} \Big],
 \eqn
where   $a_{\mu} \equiv u^{\alpha}D_{\alpha}u_{\mu}$, and $c_i$'s are arbitrary constants \footnote{Because of the hypersurface-orthogonal condition, only three of them are 
independent \cite{LACW,LGSW,Jacobson10}. But, here we shall leave this possibility open. }. Then, 
the variation of $S_\phi$ with respect to $\phi$ yields,  
 \bqn
 \lb{U6}
 D_{\mu} {\cal{A}}^{\mu}  = 0,
 \eqn
where,
 \bqn
 \lb{U7}
{\cal{A}}^{\mu} &\equiv& \frac{\left(\delta^{\mu}_{\nu}  + u^{\mu}u_{\nu}\right)}{\sqrt{X}}\AE^{\nu},\nb\\
\AE^{\nu} &\equiv& D_{\gamma} J^{\gamma\nu} + c_4 a_{\gamma} D^{\nu}u^{\gamma},\nb\\
J^{\alpha}_{\;\;\;\mu} &\equiv&  \big(c_1g^{\alpha\beta}g_{\mu\nu} +
c_2 \delta^{\alpha}_{\mu}\delta^{\beta}_{\nu}
+  c_3 \delta^{\alpha}_{\nu}\delta^{\beta}_{\mu}\nb\\
&&  ~~~ - c_4 u^{\alpha}u^{\beta} g_{\mu\nu}\big)D_{\beta}u^{\nu}.
 \eqn

To solve Eq.(\ref{U6}) in terms of $\phi$ directly, it is very complicated usually, as high-order spatial derivatives of $\phi$ are often involved, and the equation is highly nonlinear. 
So, often one divides the task into two steps: (i) One first solves it in terms of $u_{\mu}$, so the corresponding equation becomes second-order, although it is still quite nonlinear.  
(ii) Once $u_{\mu}$ is given, one can find $\phi$ by integrating out Eq.(\ref{eq1.3}). However, as far as the universal horizon is concerned, Eq.(\ref{eq1.2}) shows  that
the second step is even not needed. Therefore, to find the location of the universal horizon now reduces first to solve Eq.(\ref{U6}) to obtain $u_{\mu}$, subjected to the unit
and  hypersurface-orthoginal conditions,
\bq
\lb{HOs}
(i) \; u_{\mu} u^{\mu} = -1,\;\;\; (ii)\; u_{[\nu}D_{\alpha}u_{\beta]} = 0, 
\eq
and then solve Eq.(\ref{eq1.2}). Similar to the spherical case \cite{Jacobson10}, the four-velocity $u_{\mu} = (u_t, u_r, 0, ..., 0)$ in the spacetimes, 
\bq
\lb{metric}
ds^2  = - F(r)dt^2 + \frac{dr^2}{F(r)} + r^2 dx^idx^i,
\eq
is always hypersurface-orthoginal. Hence, the conditions given by Eq.(\ref{HOs}) in the spacetimes of Eq.(\ref{metric}) simply reduces to $u_{\mu} u^{\mu} = -1$, which can be written as
\bq
\lb{eq1.4}
u_t^2 - {\left(u^r\right)}^2 = F(r),
\eq
where $u^r \equiv Fu_r$. 

In review of the above, one can see that solving the khronon equation (\ref{U6}) now reduces to solve it in terms of $u_{\mu}$,
 subjected to the constraint (\ref{eq1.4}). As mentioned above, it is a second-order differential equation in terms of $u_{\mu}$. Therefore, to determine uniquely $u_{\mu}$, two boundary
conditions are required, which can be \cite{BS11,LACW}: (i) The  khronon vector is
aligned asymptotically with the timelike Killing vector, $u^{\mu} \propto \zeta^{\mu}$. 
(ii) The khronon field has a regular future sound horizon. 

Even with all the above simplification, it is found still  very difficult to solve khronon equation (\ref{U6})
in the general case. But, when $c_1 + c_4 = 0$ we find that  Eq.(\ref{U6}) has a
simple solution $u^r = r_B/r^d$, where $r_{B}$ is an integration
constant. Then, from Eq.(\ref{eq1.4})  we can get $u_t$, so finally we have,
 \bqn
\lb{UHB2} u^{\mu}=\delta^{\mu}_{t}
\frac{\sqrt{G(r)}}{F(r)}-\delta^{\mu}_{r}
\frac{r_B}{r^d},
 \eqn
 where
 \bq
 \lb{UHB6a}
 G(r) \equiv \frac{r_{B}^2}{r^{2d}} + F(r).
 \eq
Clearly, in order for the khronon field $\phi$ to be well-defined,  we must assume 
\bqn
\lb{2.13}
 G(r) \ge 0,
\eqn
in the whole space-time, including the internal region of a Killing horizon, in which we have
$F(r) < 0$. In addition, $u^{\mu} \rightarrow u^{t} \delta^{\mu}_{t} \propto \zeta^{\mu}$, as $r \rightarrow \infty$, as longer as $F(r = \infty)$ remains positive. The latter is true for the case where spacetimes
are either asymptotically flat or anti-de Sitter.  Moreover,  for the choice $c_{1} + c_{4} = 0$, the khronon has an infinitely large speed $c_{\phi} = \infty$ \cite{LGSW}. Then, by definition  the universal horizon 
coincides with  the sound horizon of the spin-0 khronon mode. So, the regularity
of the khronon on the  sound horizon now  becomes the regularity on the universal horizon.
On the other hand,  
from Eq.(\ref{eq1.2}) we find that
\bqn
\lb{2.14}
u_{\mu}\zeta^{\mu} &=&\sqrt{G(r)} = 0,
\eqn
at the universal horizons. Then, from the regular condition (\ref{2.13}) we can see that the universal horizon located at $r = r_{UH}$ must be also a minimum of $G(r)$. Therefore, 
at the universal horizons we must have  \cite{BBM,LACW,LGSW},
\bq
\lb{2.15}
\left. G(r) \right|_{r= r_{UH}} = 0 = \left. G'(r)\right|_{r= r_{UH}},
\eq
which are equivalent to 
 \bqn
\lb{UHB4a}
&& r_B^2 = -F(r_{UH})r_{UH}^{2d},\\
\lb{UHB4b} && 2dF(r_{UH})+r_{UH}F'(r_{UH})=0. 
\eqn 
The corresponding surface gravity  is given by \cite{CLMV},
 \bqn
 \lb{UHB6}
\kappa_{UH}&\equiv & \frac{1}{2} u^{\alpha} D_{\alpha} \left(u_{\lambda} \zeta^{\lambda}\right) \nb\\
&=&   \left.
\frac{r_B}{2\sqrt{2}r^d}\sqrt{G''\left(r\right)}\right|_{r =
r_{UH}}.
 \eqn
 
 For the solutions found in  Section III, we can see that only the generalized BTZ solutions have Killing horizons, and possibly have also universal horizons. 
 For this class of solutions, we have 
 \bq
 \lb{fr}
 F(r)=-\frac{2m}{r^{d-1}}-\frac{2\Lambda_er^2}{d(d+1)},
 \eq
for which the condition $F(r=\infty) \ge 0$ requires $\Lambda_{e} < 0$.  Applying the above formulas to this class of solutions,  we find that universal horizons indeed exist, and
are given by, 
 \bqn
 \lb{UHB7}
   r_{UH}&=&\left[-\frac{md(d+1)}{(d+2)\Lambda_e}\right]^{\frac{1}{d+1}}. 
 \eqn
 In addition, we also have 
 \bqn
 \lb{UHB7a}
 r_{EH}&=&\left[-d(d+1)\frac{m}{\Lambda_e}\right]^{\frac{1}{d+1}},\nb\\
 \kappa_{EH}&=&-\frac{\Lambda_e}{d}\left[-\frac{d(d+1)m}{\Lambda_e}\right]^{\frac{1}{d+1}},\nb\\
 \kappa_{UH}&=&\frac{m(1+d)}{\sqrt{2(2+d)}}\left[-\frac{(2+d)\Lambda_e}{d(d+1)m}\right]^{\frac{3d}{2d+2}},
 \eqn
where $r_{EH}$ and $\kappa_{EH}$ denotes, respectively, the location of the Killing horizon and the corresponding surface gravity.

\begin{figure*}[tbp]
\centering
\includegraphics[width=8cm]{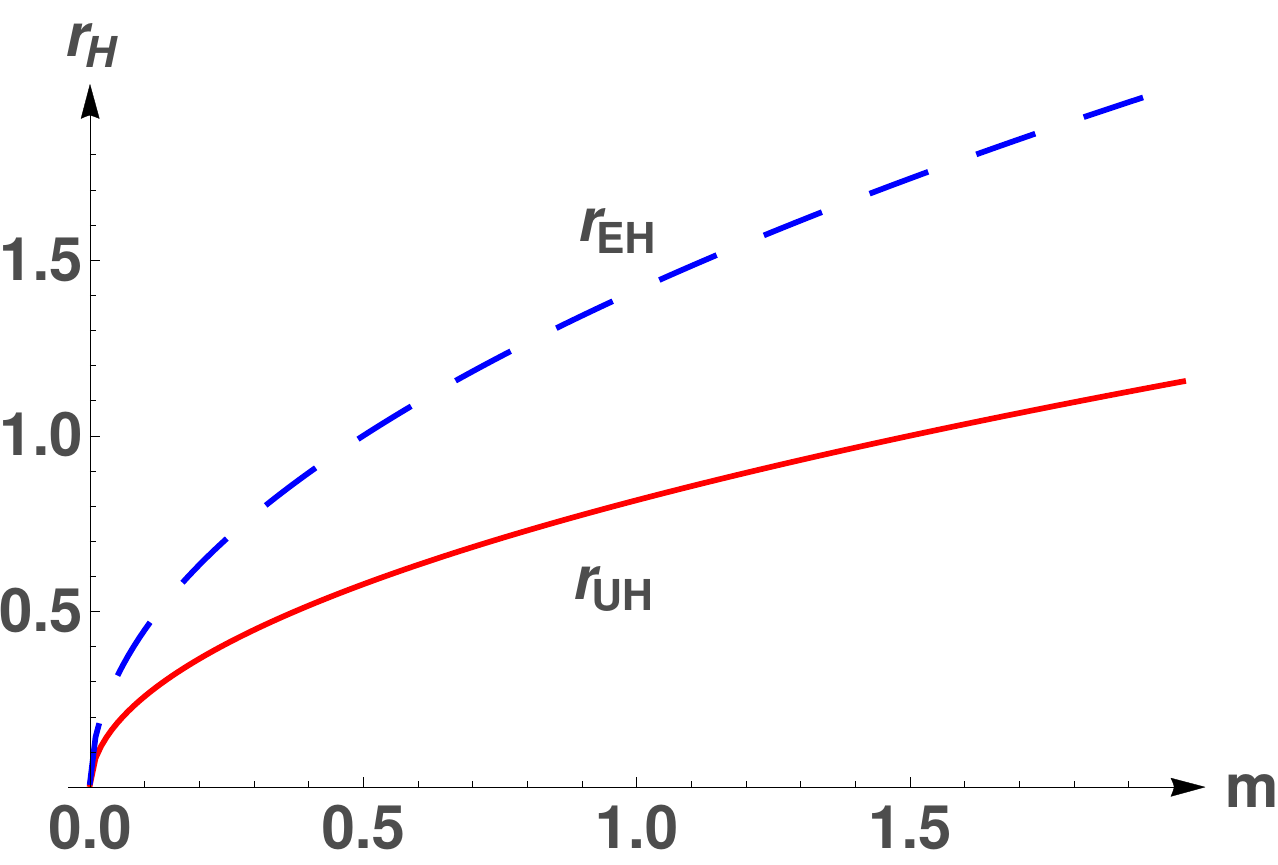}
\includegraphics[width=8cm]{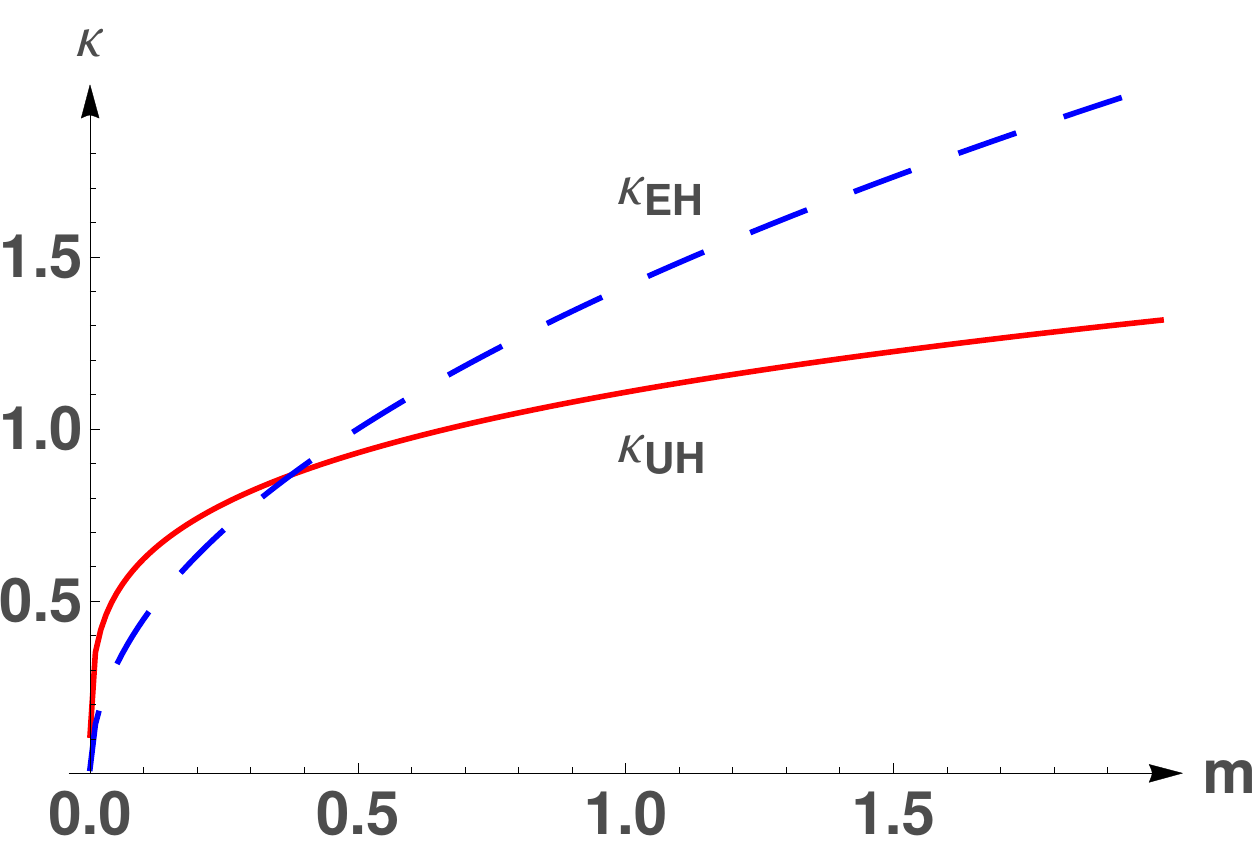}
\caption{The locations of the universal horizon $r = r_{UH}$ and
Killing (event) horizon $r = r_{EH}$ and the corresponding surface gravities $\kappa_{UH}$ and $\kappa_{EH}$ on the universal  and killing (event) horizon, respectively,  for
 the solutions $d=1$ and $\Lambda_e=-1$.}
 \label{h1}
 \end{figure*}

\begin{figure*}[tbp]
\centering
\includegraphics[width=8cm]{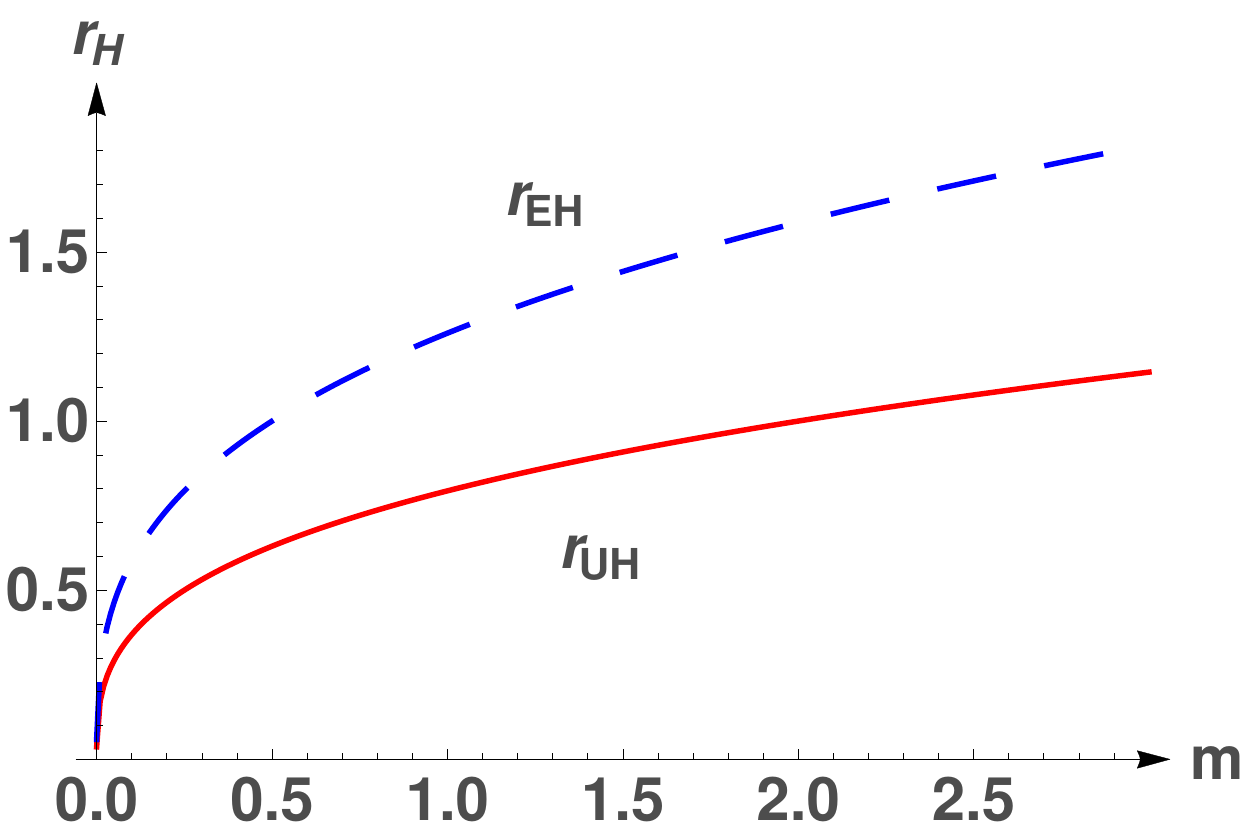}
\includegraphics[width=8cm]{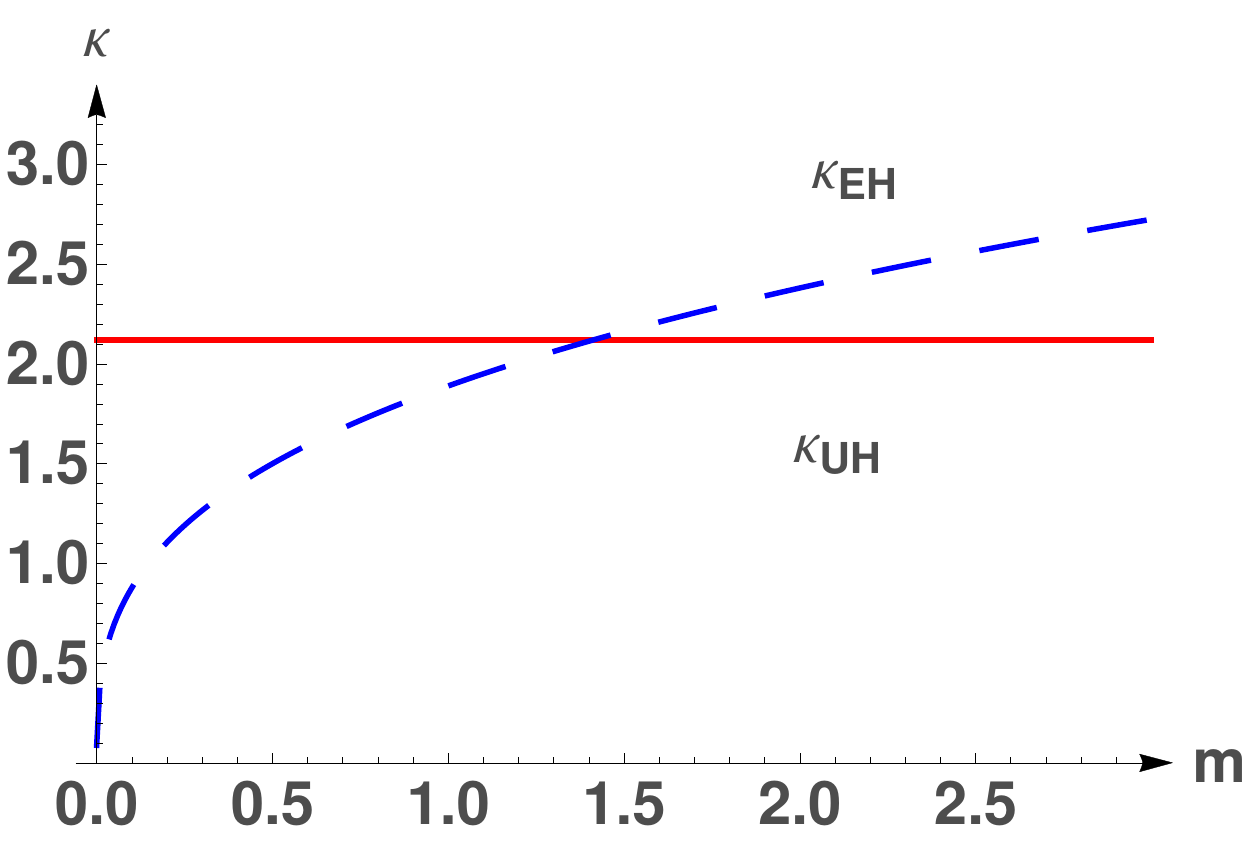}
\caption{The locations of the universal horizon $r = r_{UH}$ and
Killing (event) horizon $r = r_{EH}$ and the corresponding surface gravities $\kappa_{UH}$ and $\kappa_{EH}$ on the universal  and killing (event) horizon, respectively,  for
 the solutions $d=2$ and $ \Lambda_e=-3$. }
 \label{h2}
 \end{figure*}

\begin{figure*}[tbp]
\centering
\includegraphics[width=8cm]{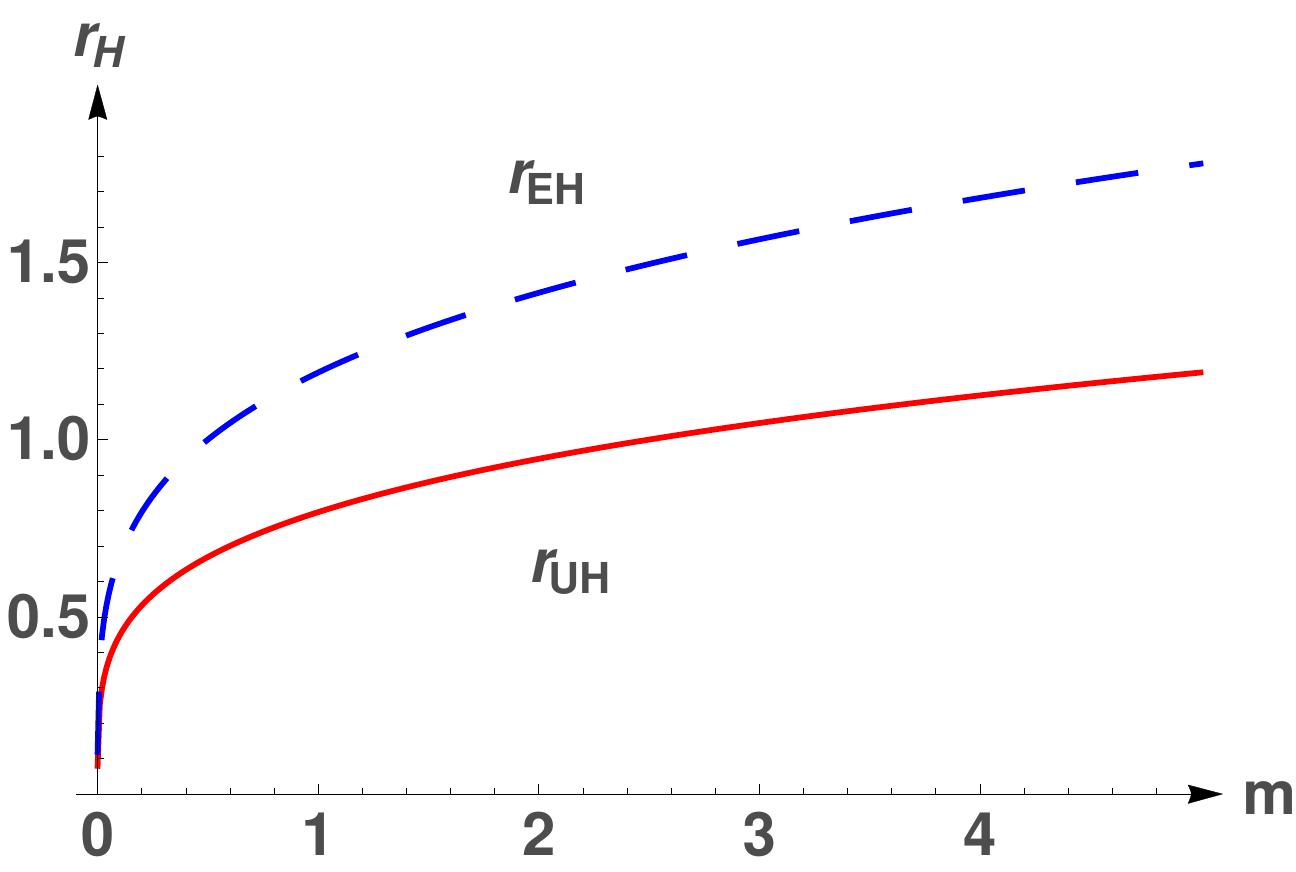}
\includegraphics[width=8cm]{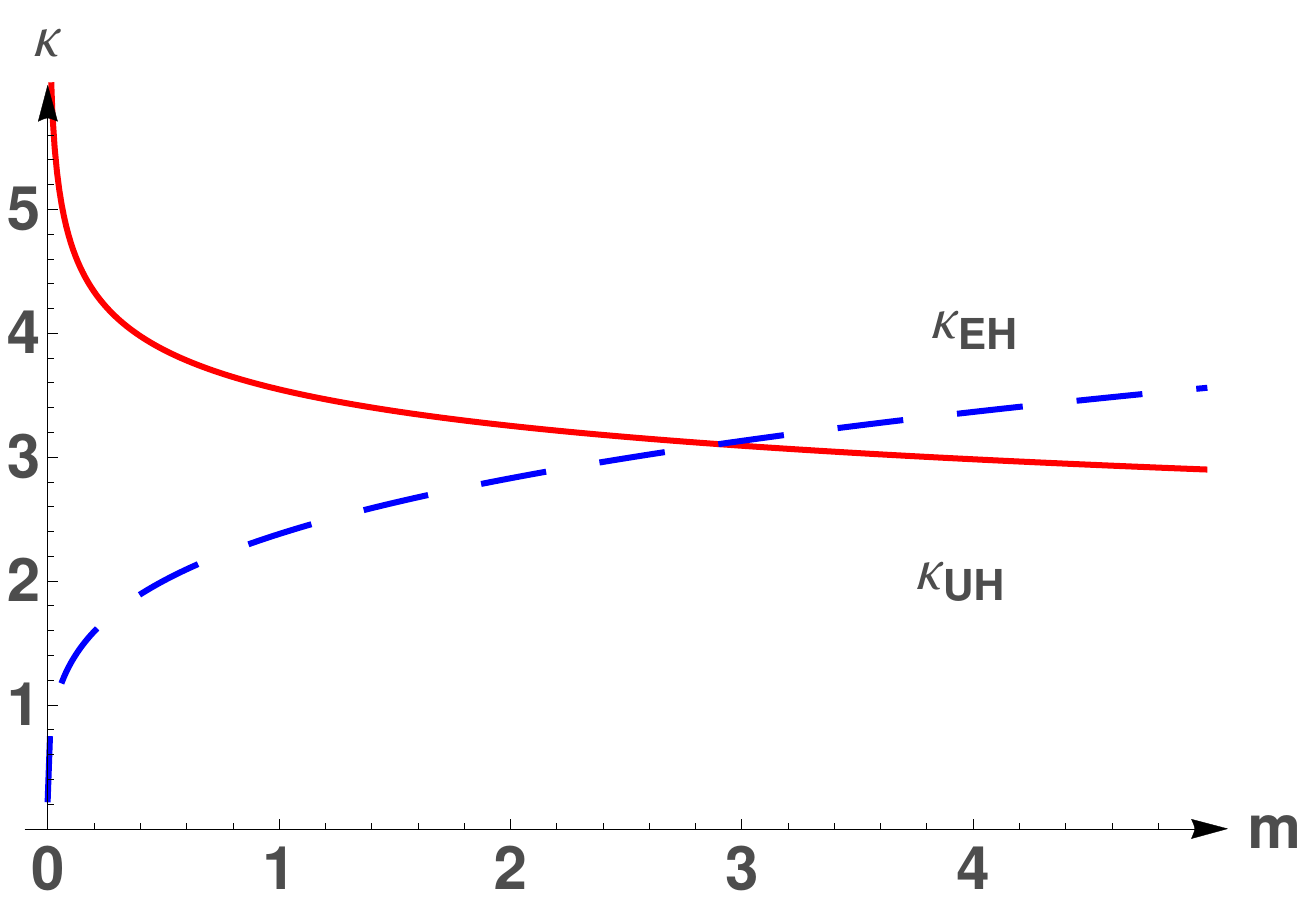}
\caption{The locations of the universal horizon $r = r_{UH}$ and
Killing (event) horizon $r = r_{EH}$ and the corresponding surface gravities $\kappa_{UH}$ and $\kappa_{EH}$ on the universal  and killing (event) horizon, respectively,  for
 the solutions $d=3$ and $ \Lambda_e=-6$.}
 \label{h3}
 \end{figure*}

Figs. \ref{h1} - \ref{h3} show   the locations   of the universal and Killing horizons vs the mass parameter $m$ in spacetimes with $d = 1, 2, 3$, respectively.
In these figures,  the corresponding surface gravities on the universal and Killing horizons  are also given. From them we can see that the 
universal horizons are always inside the Killing horizons, as they should be [cf. the  explanations given above]. On the other hand, when $m < m_{c}$, the surface gravity on the 
universal horizon is always granter than the surface gravity on the Killing horizon, where $m_{c}$ is defined by $ \kappa_{EH}(m_c) =  \kappa_{UH}(m_c)$. But, for
$m > m_c$, the opposite, i.e., $ \kappa_{EH} >  \kappa_{UH}$,  always happens. It is interesting to note that $ \kappa_{UH}$ is independent of $m$ in the case $d =2$. 
 
\section{Conclusions}
\renewcommand{\theequation}{6.\arabic{equation}} \setcounter{equation}{0}

In this paper, we have generalized our previous studies of Lifshitz-type spacetimes in the HL gravity from $(2+1)$-dimensions \cite{SLWW} to
$(d+2)$-dimensions with $d \ge 2$, and found explicitly all the static diagonal vacuum solutions of the HL gravity without the projectability condition in the IR limit.

After studying each of these solutions  in detail (in Sections III - V), we have found that these solutions have  very rich physics, and can give rise to
almost all the structures of Lifshitz-type spacetimes found so far in other theories of gravity, including  the Lifshitz spacetimes \cite{KLM,Mann},
generalized BTZ black holes \cite{BTZ},  Lifshitz solitons \cite{LSolitons},  and Lifshitz spacetimes
with hyperscaling violation \cite{GR10,OTU}, all depending on the free parameters of the solutions. Some solutions represent geodesically incomplete
spacetimes, and extensions beyond certain horizons are needed. After the extension, it is expected that some of them may represent Lifshitz-type black holes
\cite{LBHs}.

A unexpected feature is that the dynamical exponent $z$ in all the solutions can take values in the range $z \in [1, 2)$ for $d \ge 3$ and  $z \in [1, \infty)$ for $d = 2$,
because of the stability and ghost-free conditions given by Eqs.(\ref{2.20a}) and (\ref{2.20b}).  Note that in (2+1)-dimensions the range of   $z$ takes its values from the range
$z \in (-\infty, \infty)$, as shown explicitly  in \cite{SLWW}. A up bound of $z$ in high-dimensional spacetimes was also found in some numerical solutions in \cite{Mann,LSolitons,LBHs}.

Another remarkable feature is the existence of black holes in the theory, considering the fact that the Lorentz symmetry is broken in this theory and propagations  with instantaneous 
interactions exist. Similar to the Einstein-aether  theory \cite{BS11,UHs},  there exist regions that are causally disconnected from infinity by  surfaces of finite areas --- the universal 
horizons. Particles even with infinitely large velocity would just move around on these horizons  and  cannot escape to infinity.  Such charged black holes have been also found recently 
in the HL gravity \cite{LACW}. In addition, using the tunneling approach for Hawking radiation, it was shown that the universal horizon indeed radiates thermally, and a thermodynamical interpretation
of the first law is possible \cite{BBM}.
Yet,  only  the surface gravity $\kappa_{UH}$ defined by Eq.(\ref{UHB6}) is adopted, which was obtained after the nonrelativistic  nature of the particle dynamics
was taken properly  into account \cite{CLMV},   can the standard relation $T_{UH} = \kappa_{UH}/2\pi$ between the Hawking temperature $T_{UH}$ and the surface
gravity $\kappa_{UH}$ hold for the particular solutions of the Einstein-aether theory studied in \cite{BBM}.  The covariant form of  the surface gravity Eq.(\ref{UHB6}) was further confirmed 
by considering the peeling behavior of the khronon at the universal horizons for the
three well-known classical solutions, the Schwarzschild, Schwarzschild anti-de Sitter, and  Reissner-Nordstr\"om \cite{LGSW}.  It is not difficult to show that the black hole solutions presented in 
our current paper satisfy the first law of thermodynamics at the universal horizons, and the standard relation holds with the the surface gravity defined by Eq.(\ref{UHB6}).

Note that black holes  defined by  anisotropic horizons in the HL gravity were proposed recently in \cite{HMTb}, and  it would be very interesting to study space-time structures of the 
solutions presented in this paper  in  terms of these anisotropic horizons, not to mention  the  infinitely red-shifted  horizons,   proposed recently in \cite{EHs}.

With these exact vacuum solutions, it is expected that the studies of the non-relativistic Lifshitz-type gauge/gravity duality will be simplified considerably, and we
wish to return to these issues soon.  The stability of these structures is another important issue that must be addressed.

 \section*{\bf Acknowledgements}

This work is supported in part by DOE  Grant, DE-FG02-10ER41692 (A.W.);
Ci\^encia Sem Fronteiras, No. 004/2013 - DRI/CAPES (A.W.);
NSFC No. 11375153 (A.W.), No. 11173021 (A.W.),
No. 11005165  (F.W.S.);  555 Talent Project of Jiangxi Province, China (F.W.S.);
 NSFC No. 11178018 (K.L.),  No. 11375279 (K.L.);
FAPESP No. 2012/08934-0 (K.L.);
and NSFC No. 11205133 (Q.W.).

\end{document}